\documentclass[usenatbib]{mn2e}
\usepackage{graphicx,txfonts}
\def\mechanic{{\sc Mechanic}}

\def\nuoct{{$\nu$~Oct}}
\def\chip{{\sc Hipparchos}}

\def\mj{m$_{\mbox{\scriptsize Jup}}${}}
\def\Pbin{P$_{\mbox{\scriptsize bin}}${}}
\def\ms{M$_\odot${}}
\def\idm#1{\mbox{\scriptsize #1}}
\def\mvs{$\mbox{ms}^{-1}$}
\def\sun{\odot}
\def\nuoct{{$\nu$~Oct}}
\def\nuoctantis{{$\nu$~Octantis}}

\newcommand\Chi{{(\chi^2_\nu)^{1/2}}}
\usepackage{color}
\definecolor{myred}{rgb}{0.7,0.1,0.1}
\definecolor{myblue}{rgb}{0.2,0.0,0.7}
\definecolor{mybrown}{rgb}{0.5,0.2,0.0}

\newcommand\corr[1]{{\color{myred} #1}}
\renewcommand\corr[1]{{\color{black} #1}}

\newcommand\hide[1]{}
\begin{document}
%
\title[A hypothesis of the $\nu$~Oct planetary system]
{Testing a hypothesis of the $\nu$~Octantis planetary system}
%
\author[K. Go{\'z}dziewski et al.]{
Krzysztof Go{\'z}dziewski, Mariusz S{\l{}}onina, Cezary Migaszewski \& Anna
Rozenkiewicz \\ Nicolaus Copernicus University, Centre for Astronomy, Gagarin
Str. 11, PL-87-100 Toru{\'n}, Poland
}
\maketitle
%
%
\begin{abstract}
%
We investigate the orbital stability of a putative Jovian planet in a compact
binary $\nu$~Octantis reported by Ramm et al. We re-analyzed published radial
velocity data in terms of self-consistent Newtonian model and we found stable
best-fit solutions that obey observational constraints. They correspond to
retrograde orbits, in accord with an earlier hypothesis of Eberle \& Cuntz,
with apsidal lines anti-aligned with the apses of the binary. The best-fit
solutions are confined to tiny stable regions of the phase space. These
regions have a structure of the Arnold web formed by overlapping low-order
mean motion resonances and their sub-resonances. The presence of a real planet
is still questionable, because its formation would be hindered by strong
dynamical perturbations. Our numerical study makes use of a new computational
Message Passing Interface (MPI) framework \mechanic{} developed to run massive
numerical experiments on CPU clusters.
\end{abstract}
%
%
%
\section{Introduction}
%
\label{mechanic-sec-intro}
The $\nu$~Octantis is a single--line spectroscopic binary composed of the
\nuoct {}~A K1 III giant primary ($1.4\pm0.3$~\ms) and an unseen red dwarf
secondary \nuoct{}~B, K7-M1~V ($0.5\pm0.1$~\ms) separated by $\sim 2.55 \pm
0.13$~au. \cite{Colacevich1935} collected 11~radial velocities (RVs) of the
primary, and \cite{alden1939} determined the astrometric orbit. Recently, the
orbital period $P_{\idm{bin}} =1050.11\pm0.13$~days{} and eccentricity
$e_{\idm{b}} = 0.2358\pm0.0003$ were determined by \cite{ramm2009}.
Remarkably, they also derived the inclination $i_{\idm{bin}}=71^{\circ}$ with
an error less than $1^{\circ}$ on the basis of \chip{} astrometry and new 223
precision RVs. \cite{ramm2009} discovered residual variability of these RVs
at a level of 
$\sim 200$~\mvs{} and explained this by an S-type \citep{Dvorak1984} Jovian
planet\corr{,} \nuoct{}~Ab\corr{,} having a minimal mass $m_{\idm{p}}\sin i_{\idm{p}} =
2.5$~\mj{} in an orbit with a semi-major axis $a_{\idm{p}} = 1.2\pm 0.1$~au
and an eccentricity $e_{\idm{p}} = 0.123\pm0.037$.

The putative planet is unusual, because the derived semi-major axis implies
its orbit roughly in the middle between massive primary and secondary. A
formation of such a planet in \corr{the} compact binary can be hardly explained
\citep[e.g.,][]{Kley2010}. Analytic or general stability criteria formulated
for the restricted and general three-body problem like the Hill stability
criterion, the results of \cite {Holman1999}, or the resonance-overlap
criterion by \cite{Wisdom1983} are violated if the planet is assumed to be in
a prograde orbit. According with the determination of stability limits in
binaries \citep{Holman1999}, a coplanar, prograde planetary configuration
might be stable only up to astrocentric distance $\sim
0.64$~au~\citep{ramm2009}. Indeed, \cite {ec2010} found that the Keplerian
best-fit model in the discovery paper is strongly unstable in just 1--10
binary periods time-scale for the case of prograde motion. This contradicts
the planetary hypothesis, although the discovery paper basically excludes
non-planetary sources of the observed signal, like stellar spots or
pulsations.

Many explanations of the observed RVs variability are still possible, which
might resolve the apparent paradox of the strongly unstable system. The
residual RVs may be implied by stellar chromospheric activity, a different
number of planetary companions, systematic errors of the observations,
instrumental instabilities and non-Gaussian uncertainties, 
\corr{such as} the red noise
\citep{Baluev2011}, or a different orbital model than the configuration in the
discovery paper. Indeed, \cite{morais2011} consider a hierarchical triple
system with 
\corr{an unseen binary in the place of secondary} \nuoct{}~B. Such an
orbital setup forces a precession of the primary orbit around the barycentre
of the inner binary and it could mimic the RV variations attributed to the
putative planet.

Earlier, \cite{ec2010} following the stability studies of the restricted three
body problem by \cite{Jefferys1974} found that {\em a retrograde} orbit of the
planet lies in much wider stable zone than the direct configuration. In a new
work, \cite{Quarles2012} investigate this model, extending the set of orbital
configurations, through a numerical study in the framework of the elliptic,
restricted three-body problem. 
\hide{This paper refers to a classic book of
\cite{Szebehely1966}, as the source of the equations of motion
 w.r.t. the pulsating-rotating frame. In this frame, the independent
variable is {\em the true anomaly of the binary} rather than the time, as written in
\citep[][their Eqs.~1]{Quarles2012}. Obviously such
modified equations do not describe the evolution of the restricted
problem anymore, hence we consider the results in this work 
not certain.}

\corr{In more than a decade,} modeling the precision RVs teaches us that the 
dynamical analysis
of the \nuoctantis{} system done in 
\corr{the discovery paper \citep{ramm2009}, and following
studies \citep{ec2010,Quarles2012} do not seem} fully
consistent with its dynamical character. Because the secondary mass may be
almost a half of the primary mass, the relatively wide planetary orbit is
strongly perturbed. The perturbation parameter, expressed by the mass ratio of
the binary, can be as large as $\sim$0.38 \citep{ramm2009}. In such a case the
mutual interactions are important to compute the RV signal
\citep{Laughlin2001}. Furthermore, it was assumed that the whole system is
co-planar, no matter if the planetary orbit is prograde or retrograde.
However, non-planar orbits in compact binaries might appear due to violent
post--formation scenarios due to the planet--planet scattering \cite
[e.g.,][]{Adams2003}. This may imply highly inclined configurations and
their
dynamics much complex than in a coplanar case 
\citep[e.g.,][]{Migaszewski2011}. The dynamical simulations done by \cite{ec2010}
concern initially aligned orbits. \cite{Quarles2012} study a system with
\corr{initial planetary eccentricity} $e_{\idm{p}}=0$ and two initial
longitudes ${\lambda}_{\idm{p}}=0^{\circ}$,$180^{\circ}$, respectively
(corresponding to 3 o'clock and 9 o'clock positions in their terminology).
While a particular initial orientation of the orbits may be protecting factor
for the stability, the initial orbital phases should not be fixed or selected
{\em a'priori}, if the tested configuration is 
\corr{meant} to be consistent with
the observations, 
\corr{where} we aim to analyze the dynamics of a real system. 
\corr{As a simple example, consider 
Keplerian} RV signal computed from the well known formulae
$
RV(\nu) = K [ \cos(\omega + \nu) + e \cos(\omega)],
$
where $K$ is the semi-amplitude, $\omega$ is for the pericenter argument of
the orbit, $e$ is the eccentricity and $\nu$ is the true anomaly.
\corr{Function $RV(\nu)$ changes its
sign}, if instead of $\omega=0^{\circ}$ one sets $\omega=180^{\circ}$ at the
initial time. \corr{Obviously, a more complex modification of $RV(\nu)$ is introduced 
if other angles are altered, such as orbital longitudes}. In general, the initial orbital configuration determines the
dynamical evolution {\em and} the reflex motion of the primary, which should
be consistent with observations.

In this work, we aim to verify and improve the kinematic (Keplerian) model of
the \nuoct{} planetary system by searching for the best-fit configurations in
terms of self-consistent dynamical, $N$-body model \cite [][and references
therein]{Laughlin2001}. To resolve the fine structure of the
phase space, and to investigate the dynamics and long--term stability of the
planet, we apply the fast-indicator MEGNO \citep{Cincotta2003} adapted to our
new multi-CPU computing environment \mechanic {}. In this paper, we
demonstrate a non-trivial application of this software. It was announced by
\cite{Slonina2012}. 
\corr{The \mechanic{} code is described in
an accompanying work to this paper (S{\l}onina et al., in preparation), and is
freely available on-line} (\verb+http://git.ca.umk.pl+).

This paper is structured as follows. After this introduction, section~2 is
devoted to the dynamical analysis of orbital solution proposed in
\citep{ec2010} to possibly global extent. We attempt to visualize complex
resonant structures in the vicinity of presumable retrograde planetary
configurations. This part has a general character, because we analyze unstable
behaviours of S-type planets in compact binaries. In the next section~3, we
focus on an re-analysis of the RV observations and on deriving the
best-fit parameters of the \nuoct{} system in terms of three distinct models
of the RV: the Keplerian (kinematic) formulation, Newtonian (dynamic) model,
and the Newtonian model with stability constraints
\citep[GAMP,][]{Gozdziewski2008}. In section~4 we demonstrate that stable,
retrograde orbits {\em consistent with the RVs data} may be found, indeed, but
these three RV models lead to significantly different orbital architectures.
Conclusions are given in section~5.
%
%
\section{The dynamics of the \nuoctantis{} system}
%
To resolve the paradox of the \nuoctantis{} system instability, \cite {ec2010}
searched for stable orbits at a grid of primary mass ($1.1, 1.4$ and $1.7$)~
\ms, three mass ratios ($\mu$ = $0.2593, 0.2754$, $0.2908$), and 30--step
initial distance ratio $\rho$ between the planet and the secondary, $\rho \in
\langle 0.22, 0.54\rangle \equiv \langle 0.56,1.38\rangle$~au. The equations
of motion were integrated for $10^3$~years ($\sim$ 350 binary periods
$P_{\idm{bin}}$) for prograde orbits, and for $1\times10^4$ ~years ($\sim$~3500
$P_{\idm{bin}}$) for retrograde orbits. The initial orbits were {\em aligned}
with both secondaries fixed in their apoastrons with respect to the primary.
These integrations confirmed the theoretical stability limit $\rho \sim 0.25
\equiv 0.64$~au for prograde orbits in \citep{Holman1999}. For the retrograde
case, the stability limit was found much larger, indeed, $\rho \sim 0.479$
consistent with the formal $3\sigma$ best-fit error in
\citep{ramm2009}\footnote{
\corr{We found some ambiguity in the calculation of
$\rho_0$ by \cite{ec2010}, who state that} {\em in the
elliptical case, $\rho_0$ denotes the ratio of the initial distance of the
planet from the primary ($a_{\idm{p}}$) relative to the semi-major axis of the
binary components ($a_{\idm{bin}}$)}. Following this, for their retrograde
configuration stable for 10~Myr, we may compute 
$
a_{\idm{p}}(1+e_{\idm{p}}) = \rho_0 a_{\idm{bin}}$, where $\rho_{0}=0.379$,
$a_{\idm{bin}} \sim 2.55$~au, and $e_{\idm{p}} = 0.123$, hence $a_{\idm{p}}
\simeq 0.86$~au. It might be also $a_{\idm{p}} = \rho_0 a_{\idm{bin}} \simeq
0.966$~au, if the {\em initial apoastron distance} of the planet was
approximated as $a_{\idm{p}}$, or, if $a_{\idm{p}}(1+e_{\idm{p}}) = \rho_0
a_{\idm{bin}}(1+e_{\idm{bin}})$ then $a_{\idm{p}} \simeq 1.06$~au,
\corr{which is most likely approximation used in the study of \cite{ec2010}.} 
}.
%
%
\subsection{Stability of a system with initially aligned apsides}
%
%
To extend the study of \cite{ec2010}, we applied the
dynamical maps \corr{technique. It helps} to illustrate the global structure of the
phase space, as well as to identify possible resonances as the origin of
strong instability observed in this system. 
\corr{
We use MEGNO which is a measure of the maximal Lyapunov exponent 
\citep{Cincotta2003} as the stability indicator.}
Our serial-CPU software to integrate the equations of
motions and the variational equations of the planetary problem was
encapsulated in the \mechanic{} MPI module called the {\tt CSM}\footnote{
\corr{This code is available} upon request.}. A strongly
interacting planetary system with collisional orbits requires an appropriate
integrator. We applied the Bulirsh-Stoer-Gragg scheme implemented as
reliable and well tested ODEX routine \citep{odex2010}. It provides \corr{very
good accuracy} and performance \citep[see also,][]{Chambers1999}. The
total integration time of a single initial condition was not less than
$10^{4}$ periods of the binary, and in some cases more than $3\times 10^{5}$
periods. This time is typically one--two orders of magnitude longer than the
time-span of direct numerical integrations in \cite{ec2010}, and provides at
least $10$-$100$ longer estimate of the Lagrange stability, in accord 
corr{with features of MEGNO} \citep{Cincotta2003}. The time scale of MEGNO integrations is
long enough to detect most significant mean motion resonances \citep
{Gozdziewski2008}.
\begin{figure*}
\centerline{
\hbox{
   \hbox{\includegraphics[     width=00.46\textwidth]{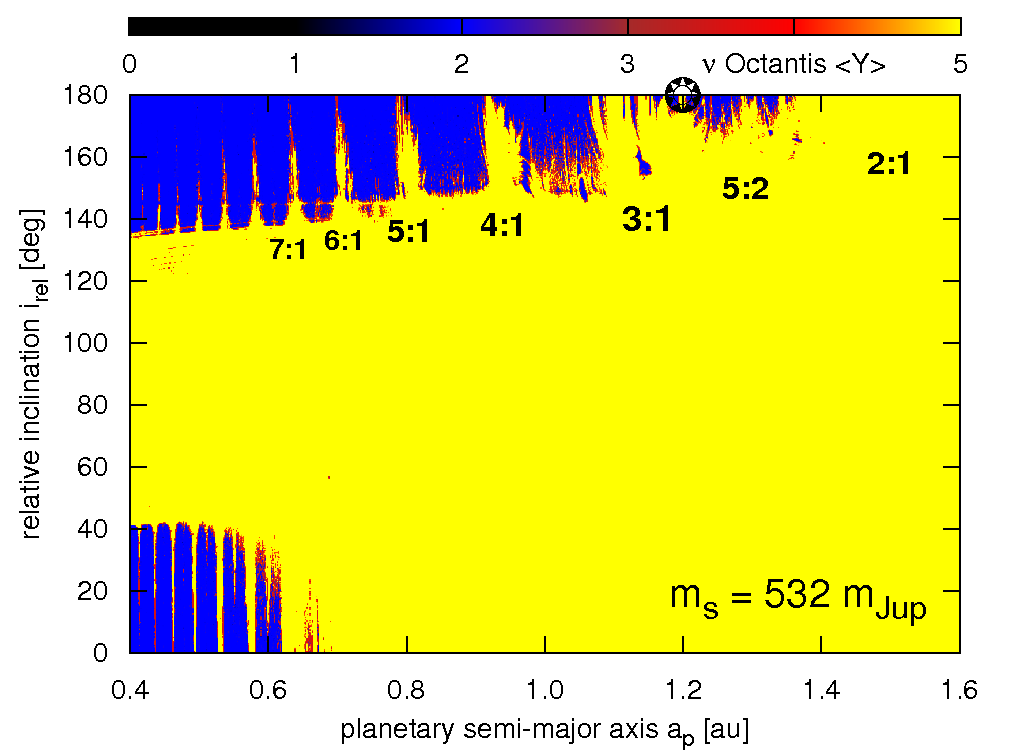}}
   \hbox{\includegraphics[     width=00.46\textwidth]{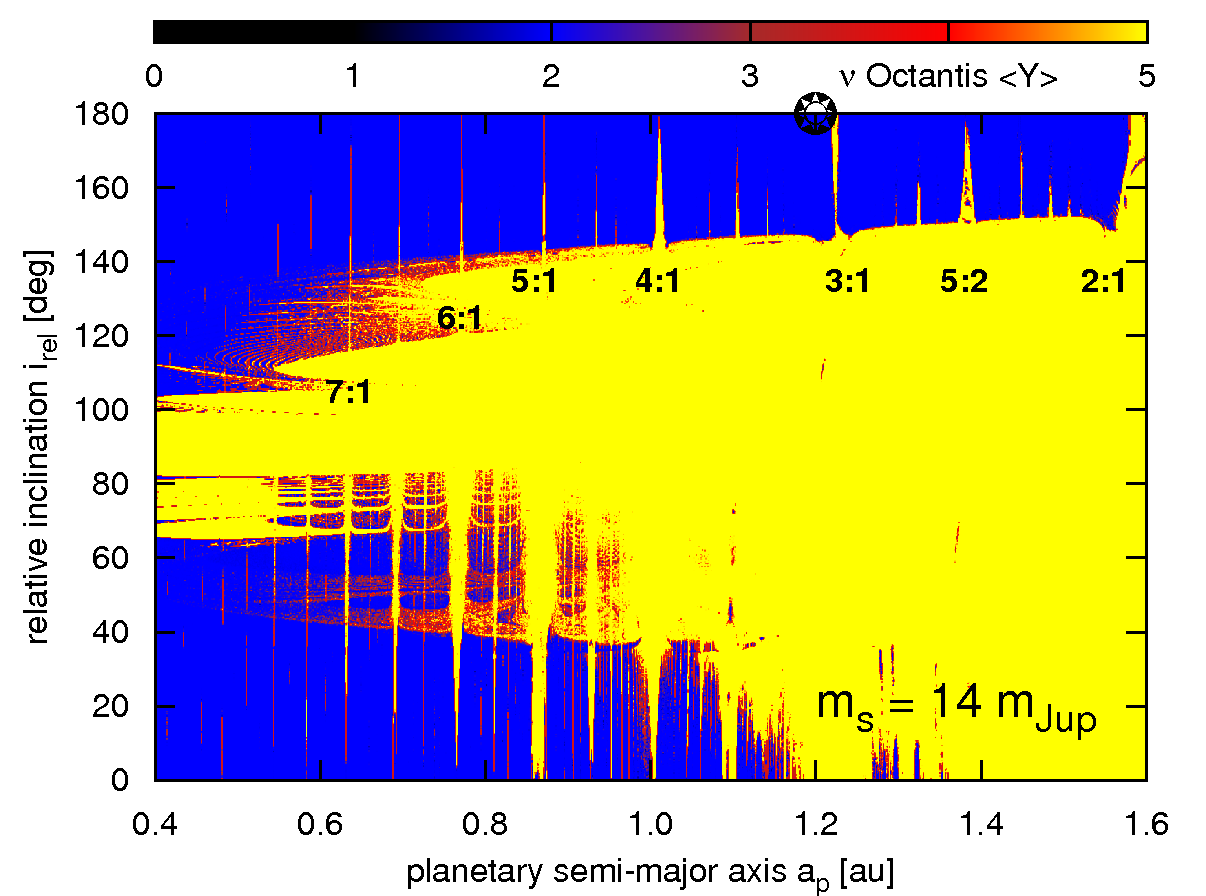}}
}
}
\caption{
MEGNO dynamical maps for the geometric setup of the \nuoctantis{} planetary
system investigated by Eberle \& Cuntz (2010). The initial apsidal lines of
the orbits are aligned. A reference position of the planet in the discovery
paper by Ramm et al. is marked with an asterisk. The low-order mean
motion resonances between the planet and the secondary are labeled. {\em The
left panel} is for the global view of the phase space for the nominal mass of
the secondary, {\em the right panel} is for an artificial brown dwarf
secondary, helpful to identify the MMRs. The raw resolution is $1440\times900$
initial conditions integrated for $\sim 3\times 10^4$ periods of the binary.
}
\label{fig:fig1}
\end{figure*}

In the first numerical experiment, we set the primary mass to $1.4$~\ms{}
which is equivalent to $\mu=0.28$. In accord with \cite{ec2010}, the planet
and the secondary are placed in the apoastrons of their orbits, hence the
initial mean anomalies ${\cal M}_{\idm{p}}={\cal M}_{\idm{S}}=180^{\circ}$, of
the planet and the secondary, respectively. Arguments of periastrons and the
nodal longitudes were set to $0^{\circ}$. In the following numerical
experiments, we varied the inclination of the planet (equivalent to the
relative inclination of the orbits) between $i_{\idm{p}}=0^{\circ}$ (direct
coplanar configuration) and $i_{\idm{p}}=180^{\circ}$ (retrograde coplanar
configuration). This setup, which we consider as {\em aligned orbits},
reproduces the initial configurations in \cite{ec2010}, i.e., the 9 o'clock
position in their terminology\footnote{In the literature, the terms of {\em
aligned} and {\em antialigned} periapses (or orbits, apsidal lines) are well
defined only for direct orbits. Here, for coplanar, direct and retrograde
orbits these terms are defined through the difference of the longitudes of
periapses, equal to $0^{\circ}$ and $180^{\circ}$, respectively.}. 
\corr{
We computed
MEGNO maps in high-resolution with a grid of $1440\times900$ pixels 
that were integrated for
25,000 periods of the binary.} The results are illustrated in
Fig.~\ref{fig:fig1}, in the semi--major axis $a_{\idm{p}}$ --- relative
inclination $i_{\idm{rel}}$--plane for the two secondary masses. 
The right map is for an
artificial brown dwarf secondary, to illustrate, through a comparison, the
strength of the perturbation in the nominal \nuoct{} system (the left panel in
Fig.~\ref{fig:fig1}). For prograde configurations, the stability limit
$a_{\idm{p}}\sim 0.6$~au is slightly smaller than its value found in the
previous papers. For retrograde orbits, it is almost doubled and reaches
$a_{\idm{p}} \sim 1.2$~au. A putative planet could be located at a border of
this zone, filled with unstable low-order MMRs. A stable retrograde orbit
found by \cite{ec2010} lies in a region spanned by strong 4:1, 3:1, 5:2 and
2:1 MMRs. A close-up of this area covering the 5:2 MMR,
\corr{slightly beyond the semi-major axis of the planet determined
in \citep{ramm2009}}, is shown in
Fig.~\ref{fig:fig2}. Overlapping of the MMRs and their sub-resonances creates
a global chaotic zone. This effect is identified as the main source of
instability of planets in compact binaries \citep {Mudryk2006}. 

The stability limit depends strongly on the relative inclination and the mass
of the secondary. Configurations with intermediate relative inclinations,
roughly between $\sim 40^{\circ}$ and $\sim 140^{\circ}$, and particularly
solutions close to polar orbits ($i_{\idm{rel}}\sim 90^{\circ}$) are very
chaotic. They are associated with the Kozai resonance \citep{Kozai1962}, see
also section 2.3. For non-coplanar orbits, the border of stable zone may be
much closer to the primary than 0.4~au. Even for a hypothetical, low--mass
secondary $\sim$14~\mj{} (Fig.~\ref{fig:fig1}, \corr{right panel}), polar orbits are very
unstable. In this case, stable prograde solutions may be found up to $\sim
1.2$~au. We also note that individual unstable MMRs are much wider for
prograde orbits than for retrograde configurations. A theoretical explanation
of this phenomenon is given in a recent paper by \cite{Morais2012}. The
stability limits are governed by the phase-space topology of retrograde and
prograde MMRs. At \corr{the} $p$/$q$ mean motion ratio, the prograde resonance is of
order $p-q$ while the retrograde resonance is of order $p+q$, hence the
resonance order is higher. Although this result is derived for the circular,
restricted three body problem, it should be valid also for
\corr{small--eccentricity binaries}. \corr{This is confirmed
by Fig.~\ref{fig:fig1} showing different widths 
of prograde and retrograde MMRs for $e_{\idm{bin}} \simeq 0.123$.}
%
%
\subsection{Developing the instability with the secondary mass}
%
%
An identification of the origin of instabilities for large mass ratio of the
binary is not straightforward. Due to overlapping MMRs and their
sub-resonances, a complex pattern of stable and unstable motions appears
(compare panels in Fig.~\ref{fig:fig1}). 
\corr{To study how the unstable zones are evolved with increasing
perturbation, we performed a numerical experiment in which the secondary
mass}
was gradually increased from an artificial value of
14~\mj{} (a brown dwarf, see the right panel in Fig.~\ref{fig:fig1}), up to
the nominal value of 532 \mj{} (the left panel of Fig.~\ref{fig:fig1}). For
the small mass secondary, the MMRs can be easily identified, and we can
follow, how they expand when the perturbation of the secondary grows. In this
way, we may also refer to a low-dimensional dynamical system given through
model Hamiltonian in \cite{Froeschle2000}. 
\corr{The dynamics of the S-type planet in the binary should reveal
qualitatively similar features because this is also governed by a perturbed
Hamiltonian}
of the form ${\cal H} = {\cal H}_0 +
\epsilon {\cal H}_1$, where ${\cal H}_0$ is the integrable (Keplerian) part,
and ${\cal H}_1$ is the perturbation due to mutual interactions of the planet
and the binary. The perturbation parameter $\epsilon$ can be expressed through 
$
\epsilon \sim m_{\idm{S}}/m_{\idm{P}} \equiv  \mu,
$ 
where $m_{\idm{S}}$ is the mass of the secondary companion $\nu$~Oct~B{} and 
$m_{\idm{P}}$ is the mass of the primary $\nu$~Oct~A{}. 

The results are illustrated through a sequence of dynamical maps in Fig.~\ref
{fig:fig3}. These MEGNO maps were computed for $\sim 10^{4}$ periods of the
binary. At each panel, we labeled the lowest-order MMRs and the mass of the
secondary (perturber). Each \mechanic{} run occupied 256~CPUs, and
computations took, depending on initial conditions (an extent of the chaotic
zone) up to a few hours of CPU time. Already for the smallest mass ratio,
corresponding to brown dwarf mass perturber, a large part of the phase space
is strongly chaotic (the top-left panel of Fig.~\ref{fig:fig3}, also the right
panel in Fig.~\ref{fig:fig1}). In the relevant zone of retrograde motions, a
pattern of narrow MMRs appears. 
A retrograde orbit might have much \corr{more} stable
space to explore than a direct orbit. In this regime, the probability of
a stable orbit would be close to~1. When the perturber mass is increased, the
widths of MMRs quickly expand, and for $m_{\idm{S}} \sim 380$~\mj{}
(roughly, the left mass limit of the secondary) one may observe 
\corr{that strong MMRs are
overlapping which emerges a zone of global chaos, roughly beyond
$a_{\idm{p}} \sim 1.2$ ~au}. For the nominal mass of the perturber (see the
bottom right panel in Fig.~\ref{fig:fig3}), there are 
\corr{only narrow} areas of stable motions. 
\begin{figure*}
\centerline{
 \hbox{
   \includegraphics[     width=00.46\textwidth]{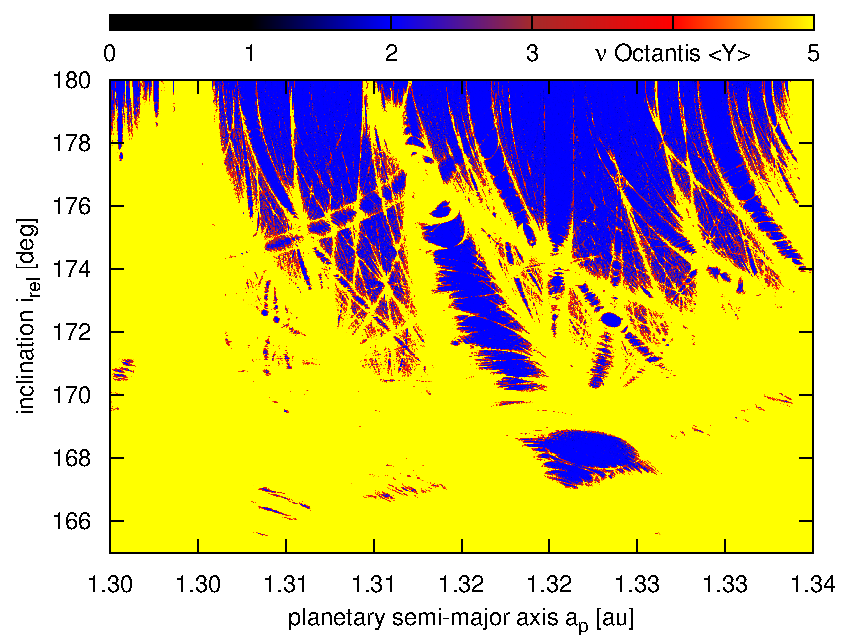}   %
   \includegraphics[     width=00.46\textwidth]{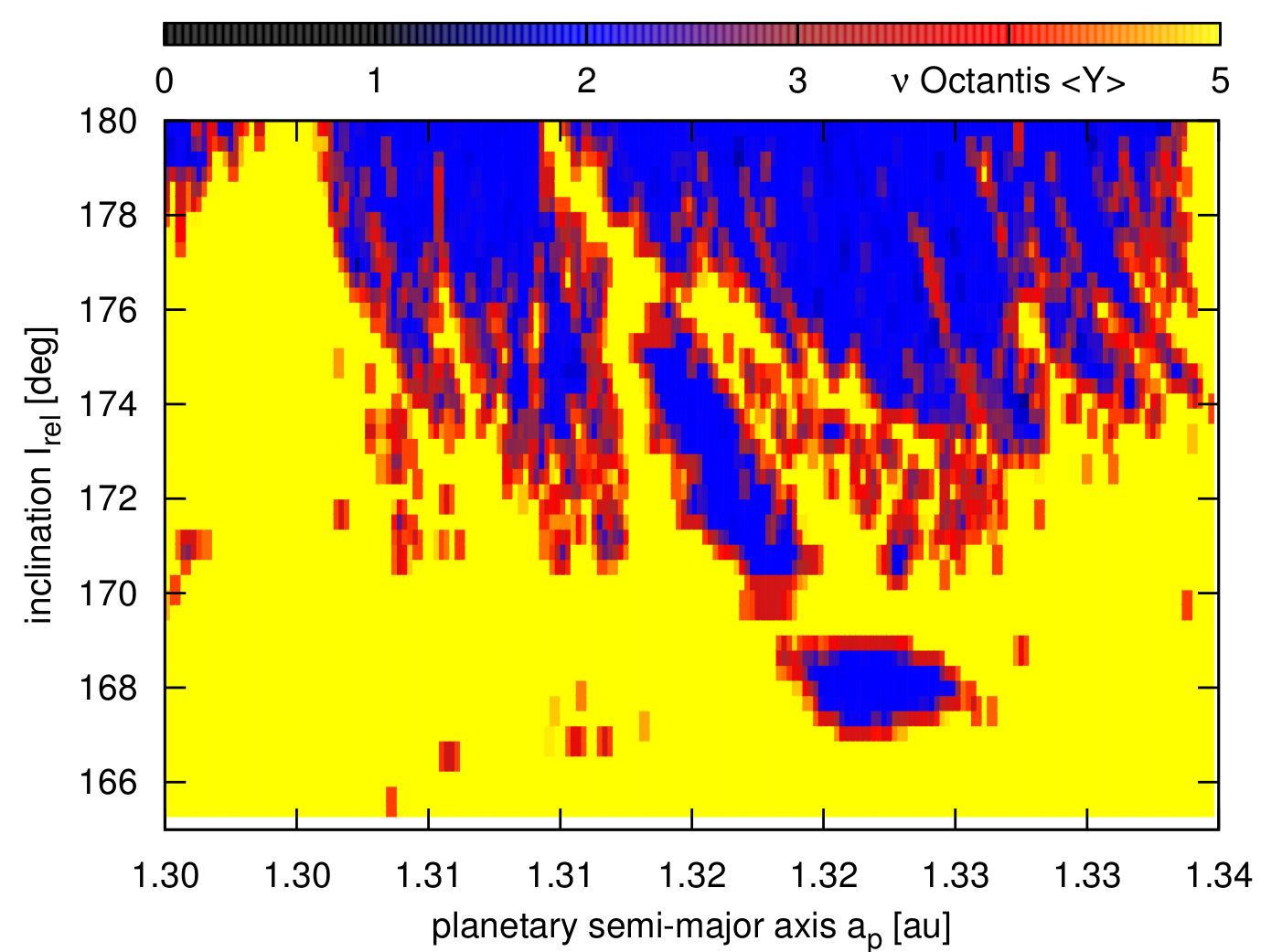}   %
}
}
\caption{
A close up of the MEGNO dynamical map for the orbital configuration with
apsidal lines initially aligned, in a neighborhood of the formal best-fit
model in (Ramm et al. 2009). The resolution of {\em the left panel} is
$1440\times900$ pixels and, for a reference, {\em the right panel} has the
resolution of $200\times50$ pixels. Each initial condition was integrated for
$\sim 3\times 10^4$ and $\sim 10^{4}$ periods of the binary, respectively.
}
\label{fig:fig2}
\end{figure*}
\corr{
A close-up} of that region in Fig.~\ref{fig:fig2} 
\corr{demonstrates} a view of the phase
space of the model Hamiltonian  for large perturbation
parameter $\epsilon$ \corr{\citep{Froeschle2000},
see also \citep{Slonina2012}}. The fine structures seen
in this map represent the Arnold web in the three body problem. They appear
due to overlapping MMRs and their sub-resonances, strongly 
\corr{split} by the
perturbation. This feature is expected \corr{because} it was found in the Outer
Solar system~\citep{Guzzo2005} and, very recently, in 
\corr{the} Kepler-11 \corr{extrasolar system 
of six planets} \citep{Migaszewski2012}. Likely, it has
not been shown in the three body problem with so much details. The right panel
of Fig.~\ref{fig:fig2} shows the MEGNO map in the same area, but with lower
resolution of $200\times50$~pixels, computed for $1 \times 10^4$ periods of the binary.
We see that a fine resolution $1440\times900$~pixels, as well as longer
integration time $\sim 3\times 10^4$~\Pbin{} are crucial to discover
\corr{the} fine details of the phase space. 
(\corr{See section~4 for} more examples of
such structures in the neighborhood of the Newtonian best-fit model of the
\nuoctantis{} system, consistent with available observations). 

The dynamical stability of planetary orbits in systems exhibiting strong
mutual interactions can be influenced even by very small changes of the
initial conditions due to the presence of unstable resonances and their
overlapping. If the mutual perturbations are small enough, and chaotic motions
appear on a regular net, they may be practically stable over very long times
\citep{Guzzo2005}. Such a state of the system is called the Nekhoroshev
regime. Otherwise, if the chaotic motions do not constitute a regular web, and
most of orbits form a global chaotic zone, the stability of the system is
influenced by a strong chaotic diffusion. This regime is related to the
resonance overlapping, and is called the Chirikov regime \citep
{Froeschle2000,Guzzo2005}. 

The Arnold web emerging due to three--body and four--body MMRs was
investigated by \cite {Guzzo2005,Guzzo2006} in the Outer Solar system. He
mapped the phase space in the semi-major axes planes of Jupiter, Saturn,
Uranus and Neptune. Here, we illustrate this feature in the three--body
problem in the semi-major axis --- inclination plane, corresponding to a
different choice of the canonical actions, and appearing due to {\em
two--body} MMRs. The Arnold web creates an intermediate zone between the
ordered and strongly chaotic motions, which spans relatively wide range of the
inclination, extending for $10^{\circ}$. In this intermediate region, the
phase-space motions which are chaotic, may persist over long periods of time. 
The results of this experiment also show that the stability limits for
co-planar binaries derived numerically by \cite{Holman1999} may be strongly
affected by the inclination. In our case of moderate eccentricity of the
planetary orbit, these limits are roughly valid for $i_{\idm{rel}} \in
(0^{\circ},40^{\circ})$ and $i_{\idm{rel}} \in (140^{\circ},180^{\circ})$. A
determination of stability limits in systems with non-zero relative
inclinations needs \corr{additional} extensive numerical study recalling that our
computations concern a particular initial configuration of the orbits. We
postpone this subject to a new work.
\begin{figure*}
\centerline{
\vbox{
 \hbox{
   \includegraphics[     width=00.44\textwidth]{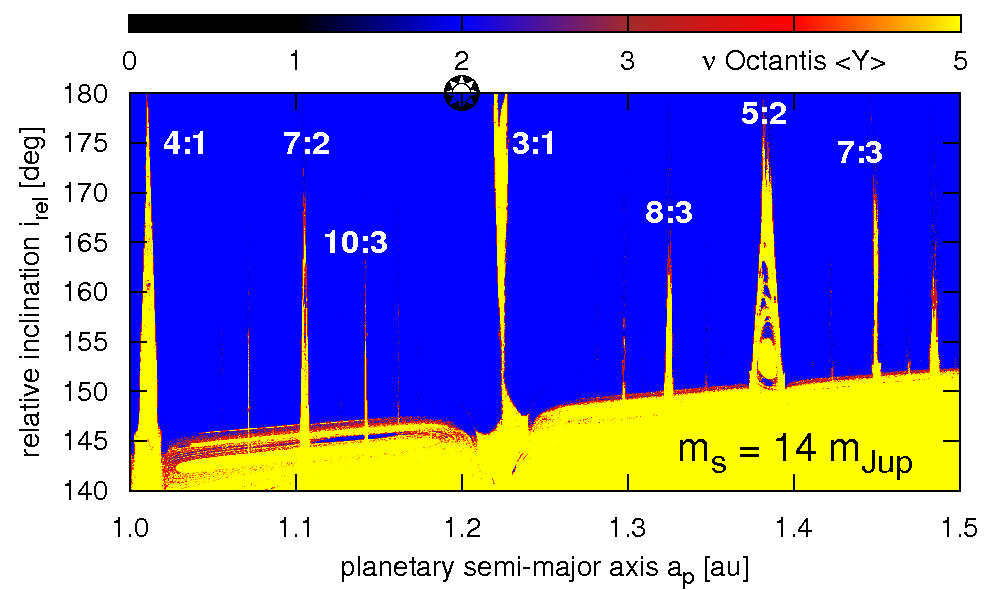} 
   \includegraphics[     width=00.44\textwidth]{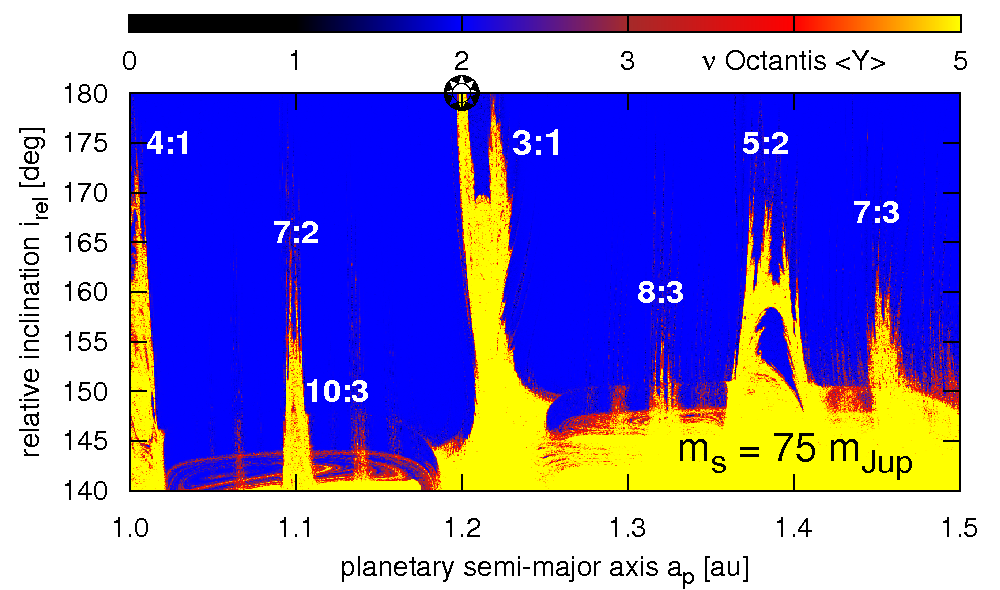} 
 }
 \hbox{
   \includegraphics[     width=00.44\textwidth]{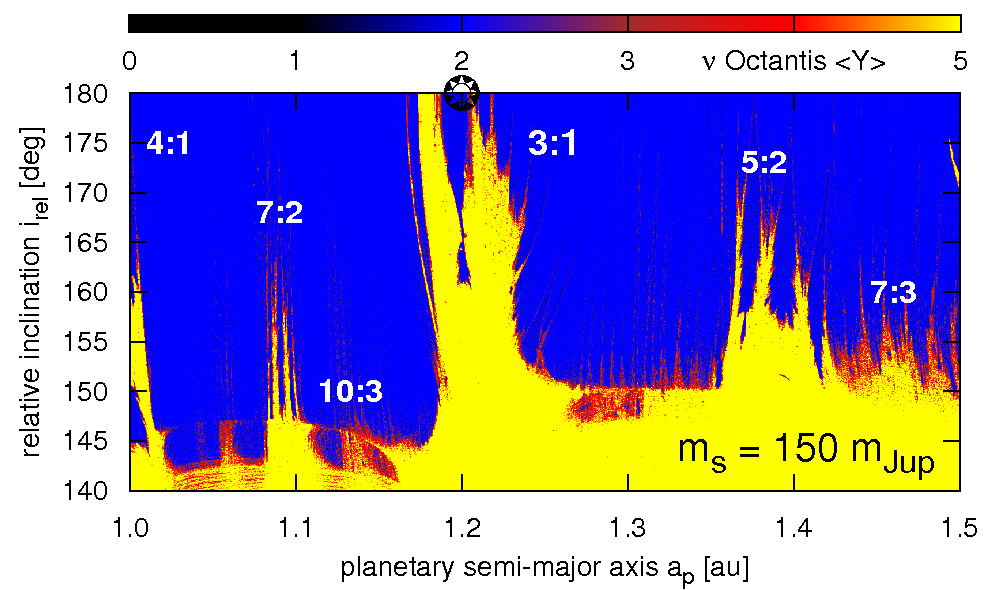} 
   \includegraphics[     width=00.44\textwidth]{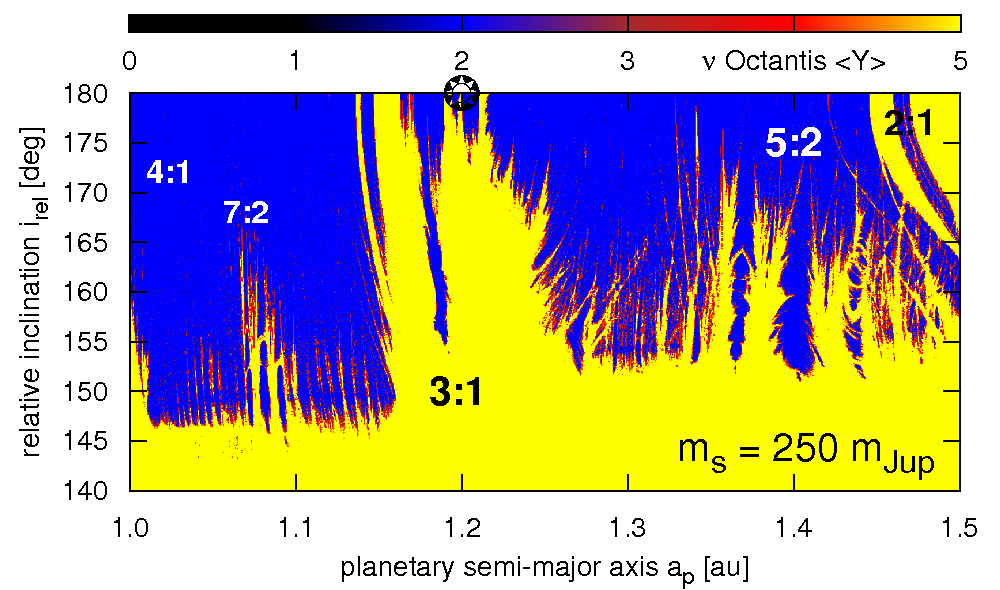} 
 }
\hbox{
   \includegraphics[     width=00.44\textwidth]{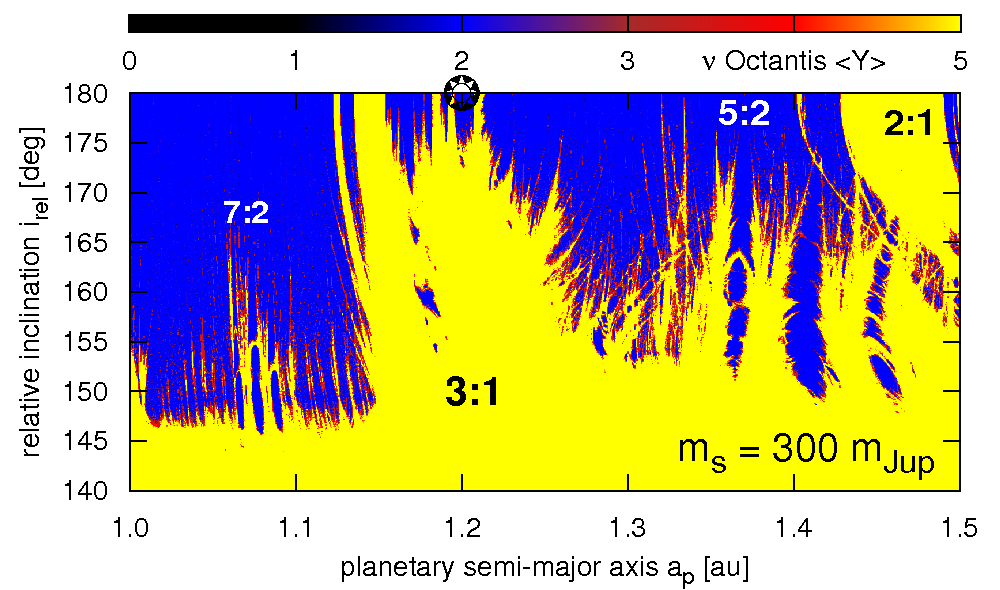}
   \includegraphics[     width=00.44\textwidth]{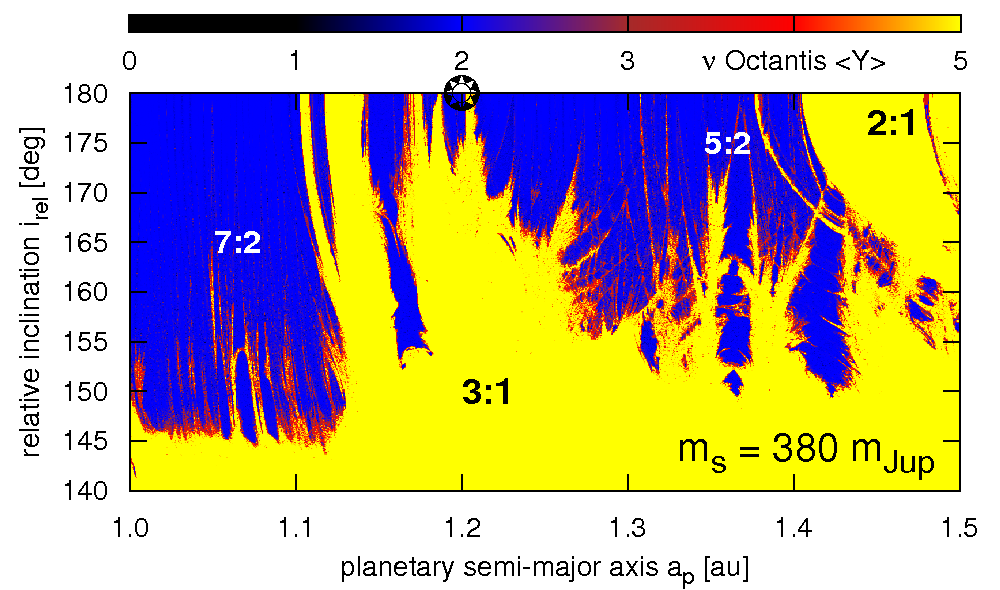} 
 }
\hbox{
   \includegraphics[     width=00.44\textwidth]{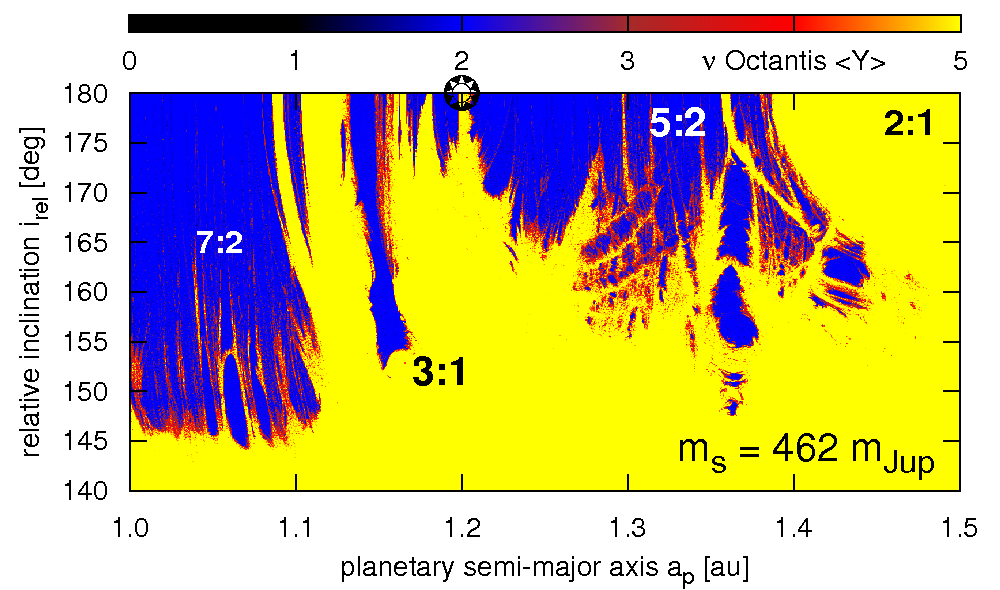} 
   \includegraphics[     width=00.44\textwidth]{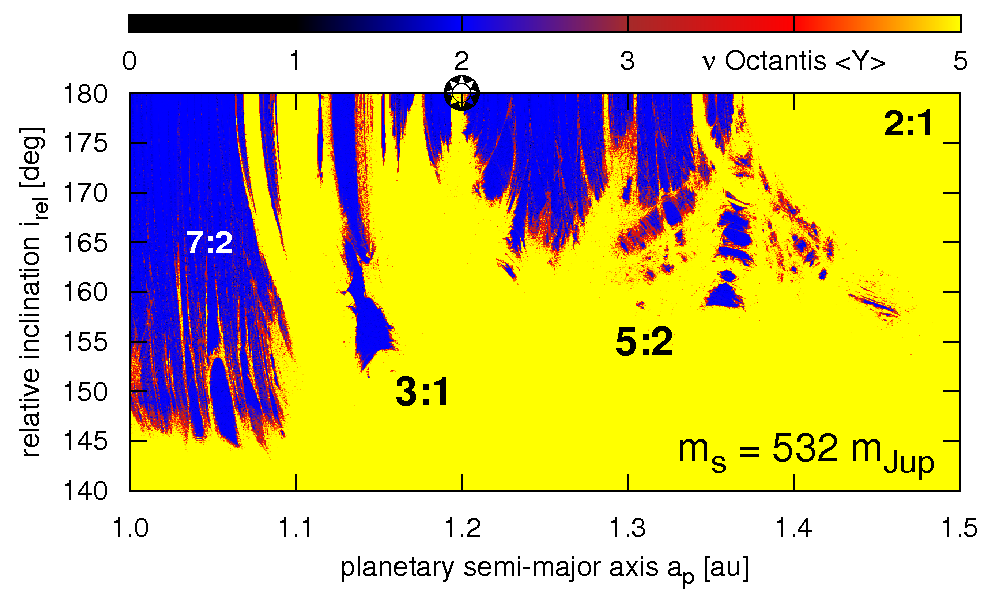} 
 }
}
}
\caption{
The results of a numerical experiment in which the secondary mass was
gradually increased from 14~\mj{} ({\em the top left panel}), through 75~\mj,
150~\mj, 250~\mj, 300~\mj, 380~\mj, 462~~\mj, up to the nominal mass 532~\mj{}
of the \nuoctantis{}B ({\em the bottom right panel}). Initial geometric
configuration is fixed the same as in the retrograde orbit model by Eberle \&
Cuntz (2010). A reference position of the planet is marked with an asterisk.
Most significant MMRs between the planet and the secondary are labeled. The
MEGNO integration time is $\sim 1\times 10^4$ periods of the binary.
}
\label{fig:fig3}
\end{figure*}
%
%
\subsection{The secular dynamics of the planet}
%
%
The origin of a wide strip of unstable motions around polar orbits can be
explained with the help of the secular octupole--level theory \citep
[e.g.,][and references therein]{Migaszewski2011}. For a direct comparison with
the MEGNO integrations, we focus again on the initial orbital setup in
\citep{ec2010}. For small $a_{\idm{p}}\sim 0.6$~au and less, the expansion
parameter $\alpha = a_{\idm{p}}/a_{\idm{bin}} \sim 0.25$ 
\corr{where} the system may be
considered as non-resonant. This permit us to study the long--term dynamics
through the secular theory. The range of $a_{\idm{p}} \in [0.4,1.2]$~au is
investigated
\corr{
to compare both relevant intervals of direct and
retrograde motion
}
although $\alpha$ becomes as large as $0.5$ at the right border
of this interval \corr{and the assumptions of the
secular theory may be questionable}. 

\begin{figure*}
\centerline{
\vbox{
 \hbox{
   \includegraphics[width=0.42\textwidth]{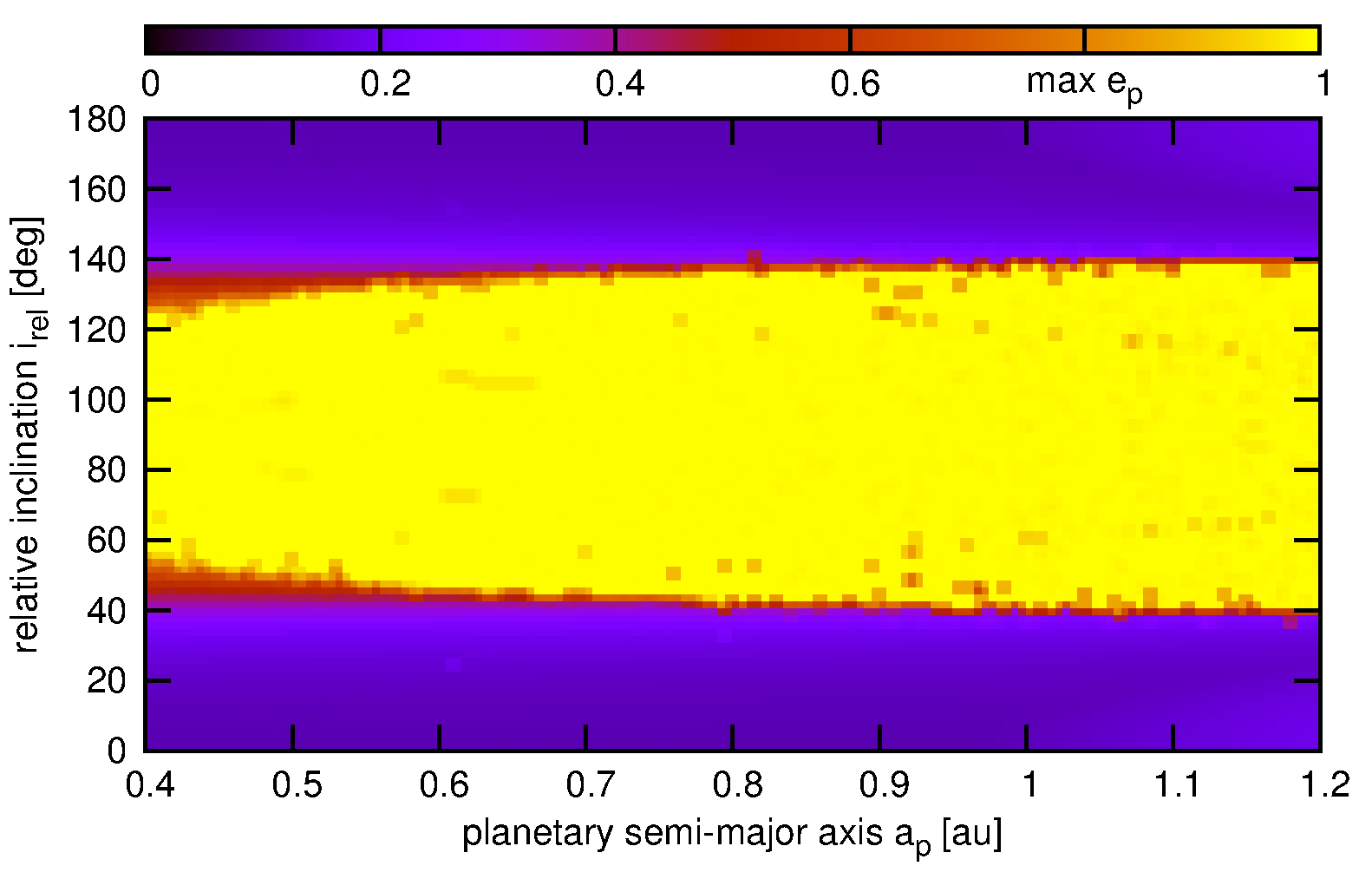} 
   \includegraphics[width=0.42\textwidth]{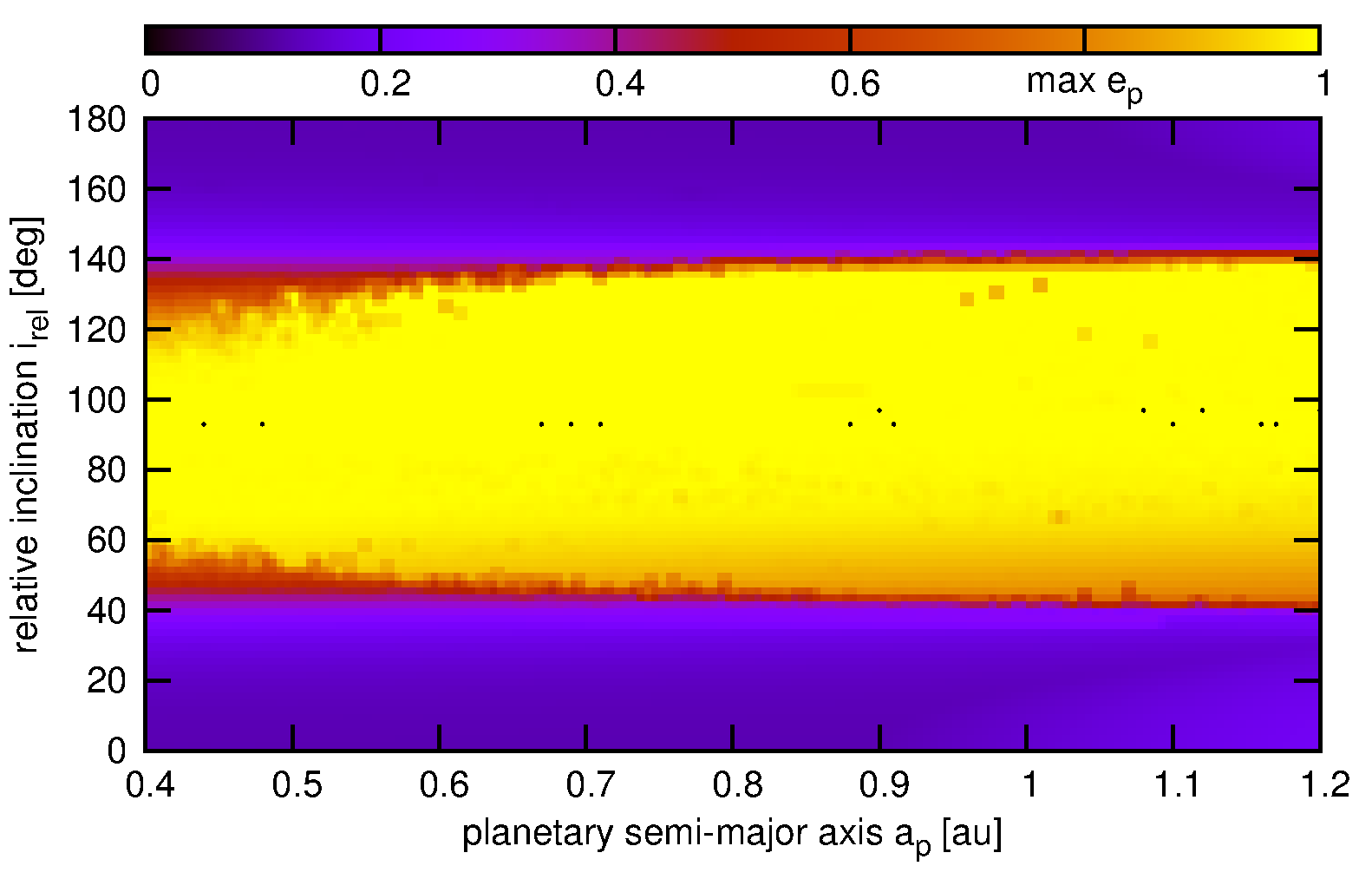} 
 }
 \hbox{
   \includegraphics[width=0.42\textwidth]{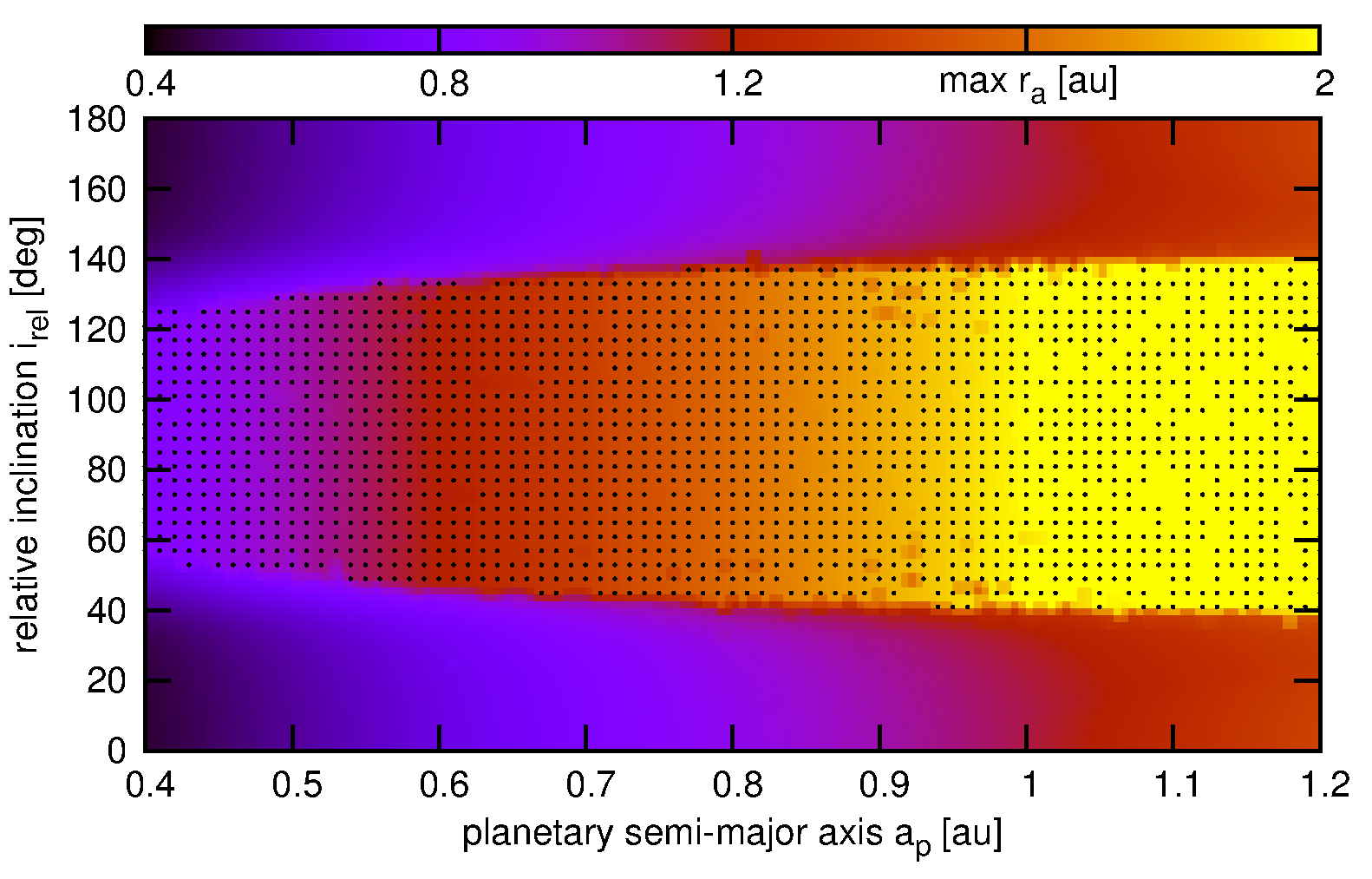}
   \includegraphics[width=0.42\textwidth]{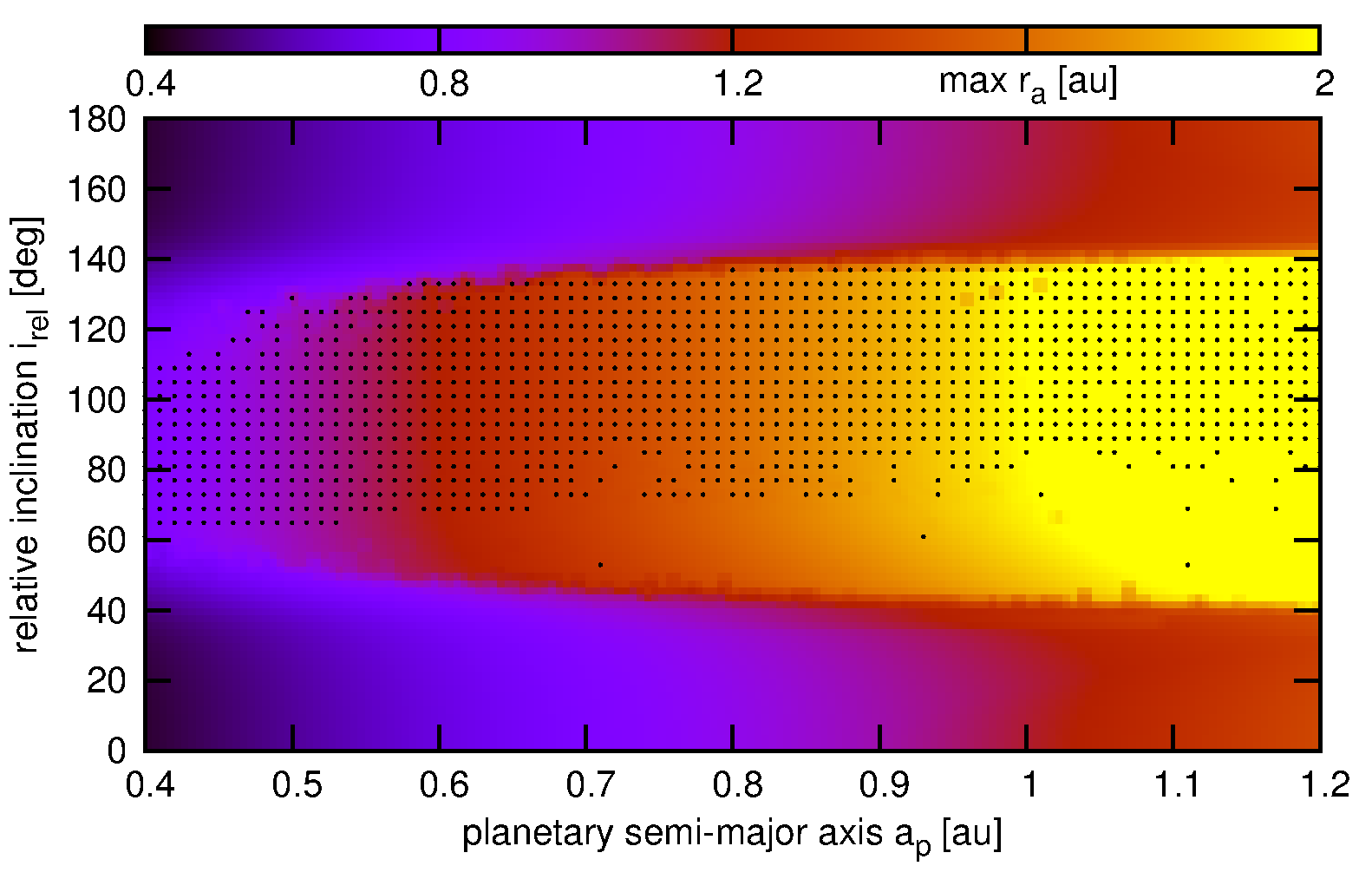} 
 }
\hbox{
   \includegraphics[width=0.42\textwidth]{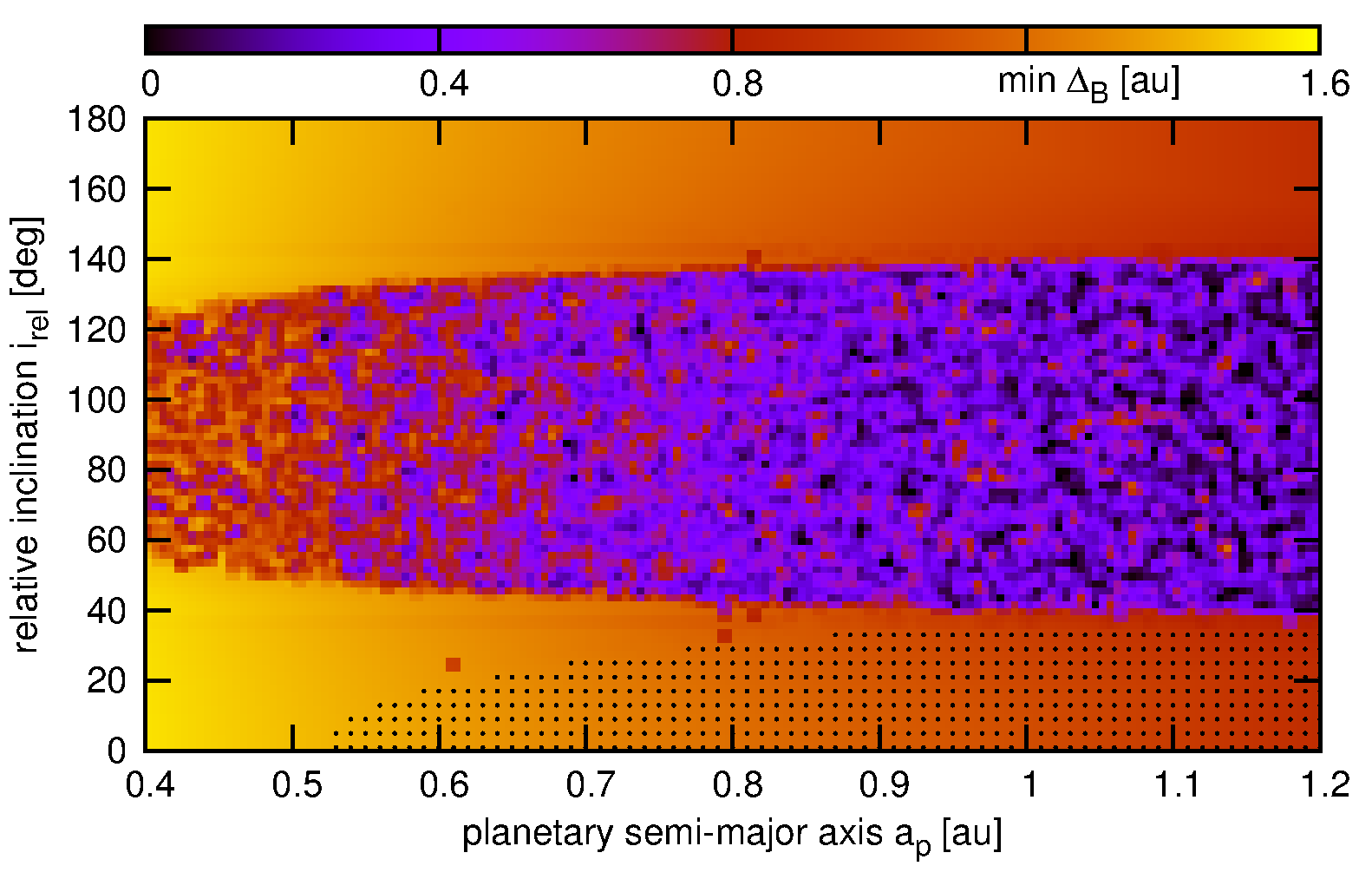} 
   \includegraphics[width=0.42\textwidth]{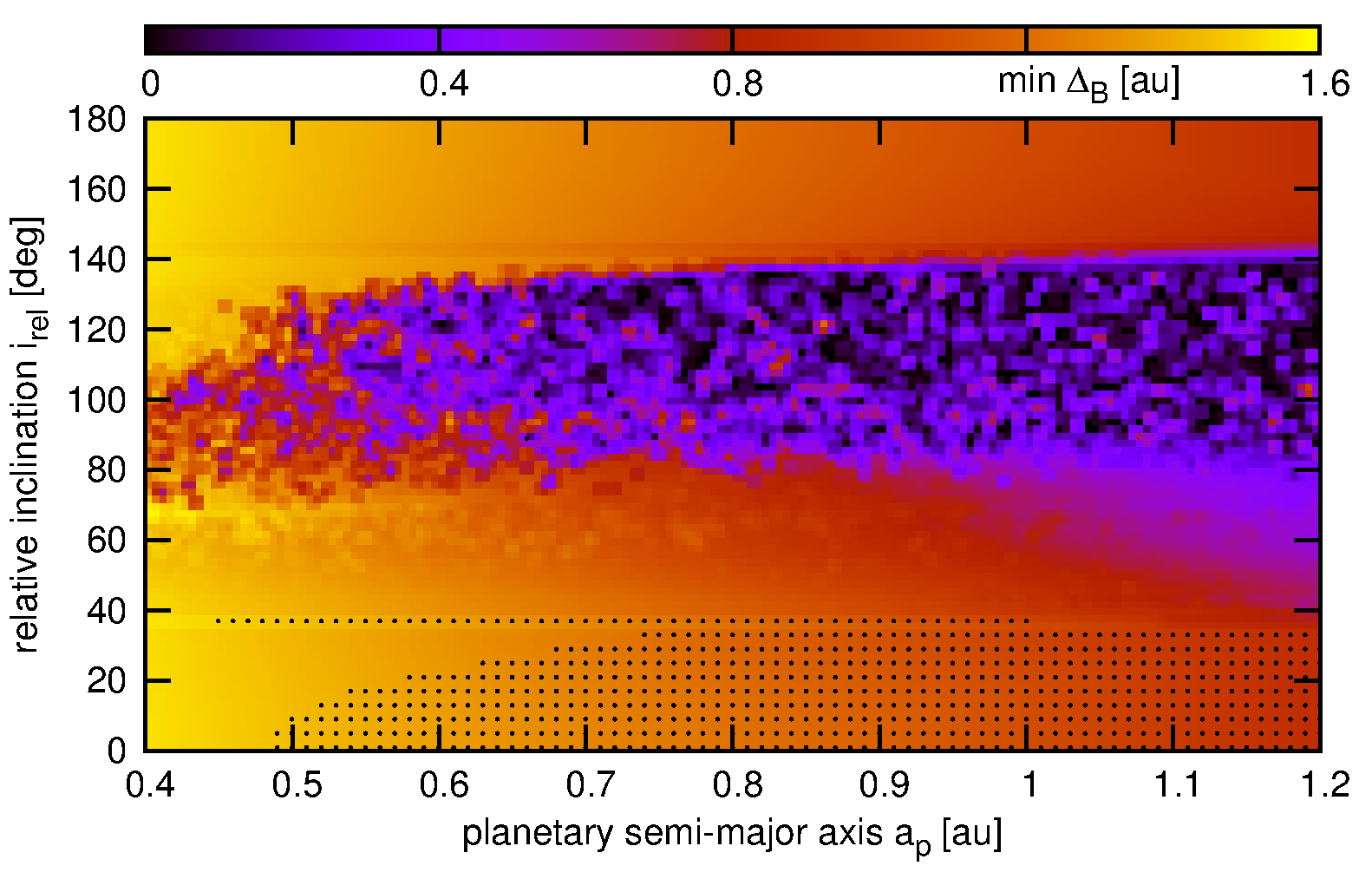} 
 }
}
}
\caption{
The orbital evolution of the secular system in terms of the octupole--level
expansion w.r.t. $\alpha=a_{\idm{p}}/a_{\idm{bin}}$. {\em The left column} is
for the nominal secondary $m_{\idm{B}}=$532~\mj{}, {\em the right column} is
for the secondary mass $m_{\idm{B}}=$14~\mj{}. The initial geometric elements
are fixed to keep $\Delta\varpi \equiv
\varpi_{\idm{p}}-\varpi_{\idm{S}}=0^{\circ}$ in the Laplace reference frame
(aligned orbits). {\em The top row} is for the maximum eccentricity of the
planet, {\em the middle row} is for the maximum of the apocenter distance of
the planet and { \em the bottom row} is the minimum distance of the planet
from the perturber. Dotted areas mark librations of $\Delta\varpi$ in the top
panels, collisional orbits with the primary in the middle panels, and
librations of $\Delta\varpi$ in the bottom panels.
}
\label{fig:fig4}
\end{figure*}

The results are illustrated in Fig.~\ref{fig:fig4} in the form of dynamical
maps in the $(a_{\idm{p}},i_{\idm{rel}}$)--plane. The left column is for the
nominal mass of \nuoct{}~B, and the right column is for an artificial,
low-mass secondary of $14$~\mj{}. The top row illustrates the maximal
eccentricity attained by the planet over the secular orbital period. Both
perturbers are massive enough to force extreme eccentricity $\sim 1$. This
must imply collisions with the primaries. Indeed, in the middle panels showing
the maximum apocenter distance of the planet, the filled dots mark initial
conditions that led to a collision with the star. These two zones closely
coincide. \corr{If the initial semi-major axis is increased}, the minimum
distance of the planet to the secondary becomes almost always close to 0~au
(see the bottom row in Fig.~\ref{fig:fig4}). Filled dots in these panels mark
regions, in which angle $\Delta\varpi\equiv \varpi_{\idm{p}}-\varpi_{\idm{S}}$
librates. 

This experiment reveals that a continuous unstable zone between $i_{\idm{rel}}
\in (40^{\circ},140^{\circ})$ predicted by the secular model closely coincides
with unstable region around the polar orbits exhibited in the full three-body
system. It means that this unstable region appear due to the Kozai resonance
which is able to force large eccentricities in the secular time-scale. More details and a
concise description of the Kozai resonance in the context of planetary
dynamics may be found in a recent review  \cite[][p.~1065--1068, and
references therein]{Beauge2012}. Hence the primary source of instability can
be identified with {\em a secular effect} rather than with short-term chaotic
dynamics associated with the MMRs. Outside unstable zones of the Kozai
resonance, the primary source of instability are the MMRs which appear as
discrete areas of chaos for small value of the perturbation parameters, and a
zone of global chaos when the perturbation becomes sufficiently strong and the
short-term MMRs may govern the global dynamics. Let us note, that
although the global picture of the
secular system is illustrated here for the aligned 
configuration, it remains qualitatively the same for the anti-aligned
orbits.
%
%
\section{Re-analysis of the RV data}
%
It is well known that the RV measurements errors originate from the formal
 errors $\sigma_{\idm{m}}$ of the Doppler method, systematic uncertainties due
 to instrumental instabilities, as well as from internal chromospheric
 variability $\sigma_{\idm{jitter}}$ of the parent star. The later error is
 difficult to quantify particularly for young or active stars. In the recent
 literature, it is common to derive astrophysical estimates of the jitter
 $\sigma_{\idm{jitter}}$ \citep[][]{Wright2005}. Joint error of the
 measurements are obtained by adding both uncertainties in quadrature for each
 data point. Recently, \cite{Baluev2009} has shown how to estimate
 $\sigma_{\idm{jitter}}$ as an additional {\em free} parameter of the fit
 model. Here, we follow the earlier, somehow more {traditional} approach. In
 this work, we computed the best-fit models after rescaling the measurement
 errors in \citep{ramm2009} by two values of $\sigma_{\idm{jitter}}$. For the
 first class of models, we \corr{choose} $\sigma_{\idm{jitter}}=5$~\mvs{}, which is
 relatively small value, exhibited by quiet, non-active stars, and likely too
 conservative in our case. The second family of solutions was obtained for
 $\sigma_{\idm{jitter}}=20$~\mvs{}. This value of jitter uncertainty seems
 better justified recalling large residuals of the Keplerian model in the
 discovery paper of a similar magnitude ($\sim 20$~\mvs{}). 
Using these RV data, we optimized different 2-planet models. The
astrometric data gathered by \chip{} were used indirectly, by fixing the nodal
line and inclination of the binary orbit close to the combined astrometric and
radial velocity solution in \citep{ramm2009}. 
\subsection{Keplerian (kinematic) model of the RV}
In the first fitting experiment, we search for the best-fit Keplerian,
non-interacting orbits. Hence, we basically follow the discovery paper. The
best-fit configurations of \nuoct{} system were found after extensive search
performed with the hybrid optimization algorithm \citep[see][and references
therein]{Gozdziewski2008}. It consists on two stages: a quasi-global Genetic
Algorithm which provides many reasonably precise models, and the
Levenberg-Marquardt (LM) method which is used to refine these solutions. 

\begin{figure*}
\centerline{
\vbox{
 \hbox{
   \includegraphics[width=0.42\textwidth]{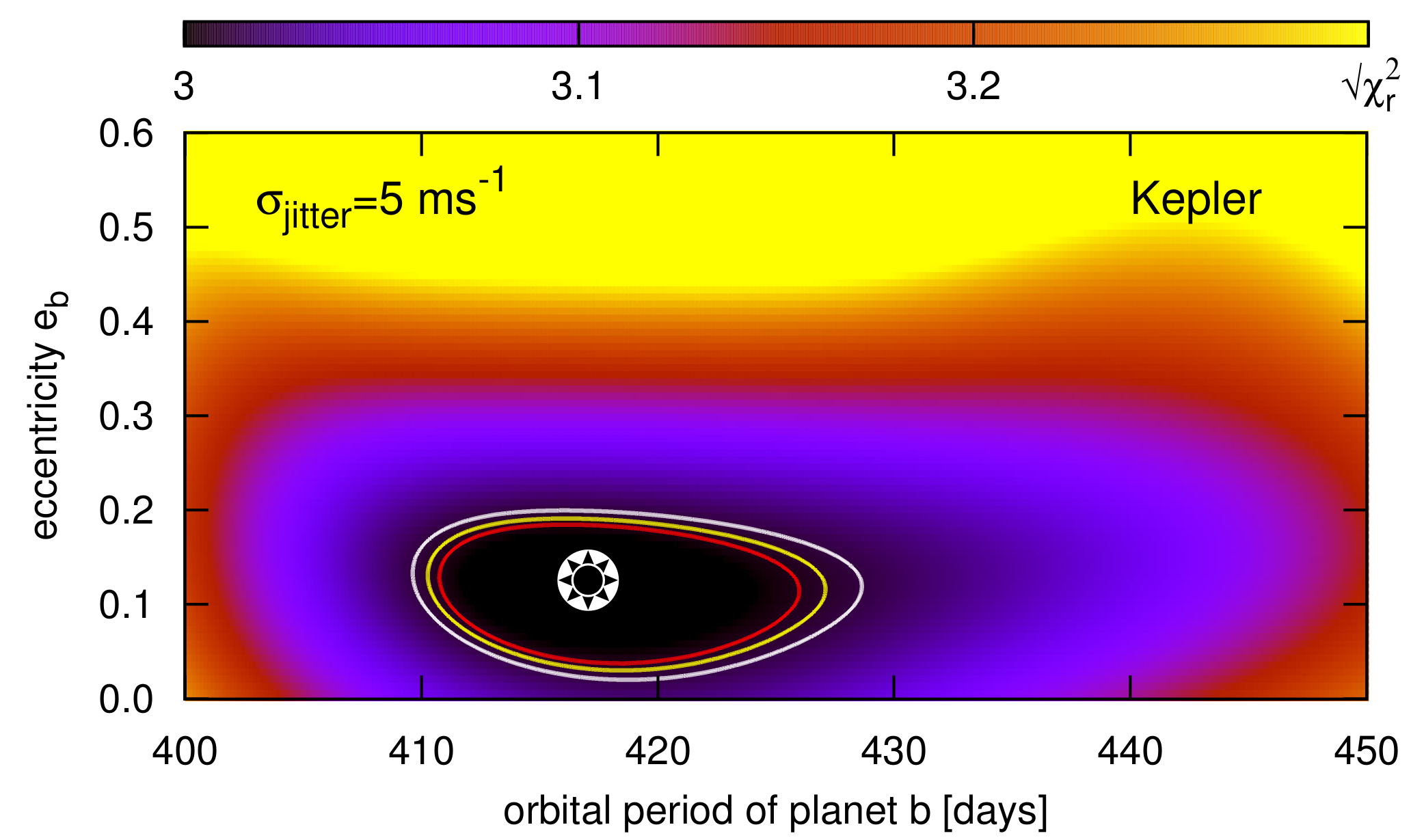} 
   \includegraphics[width=0.42\textwidth]{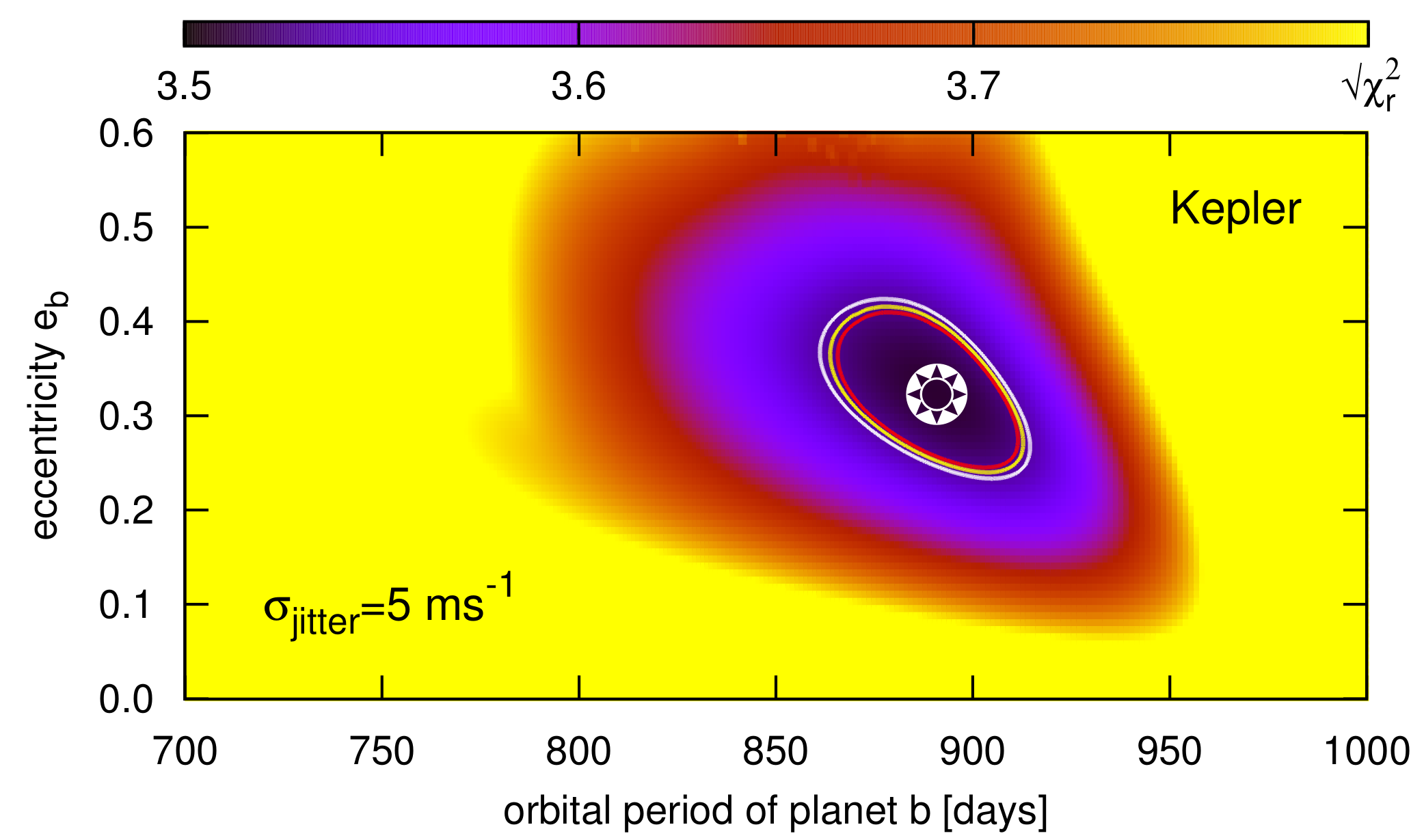} 
 }
 \hbox{
   \includegraphics[width=0.42\textwidth]{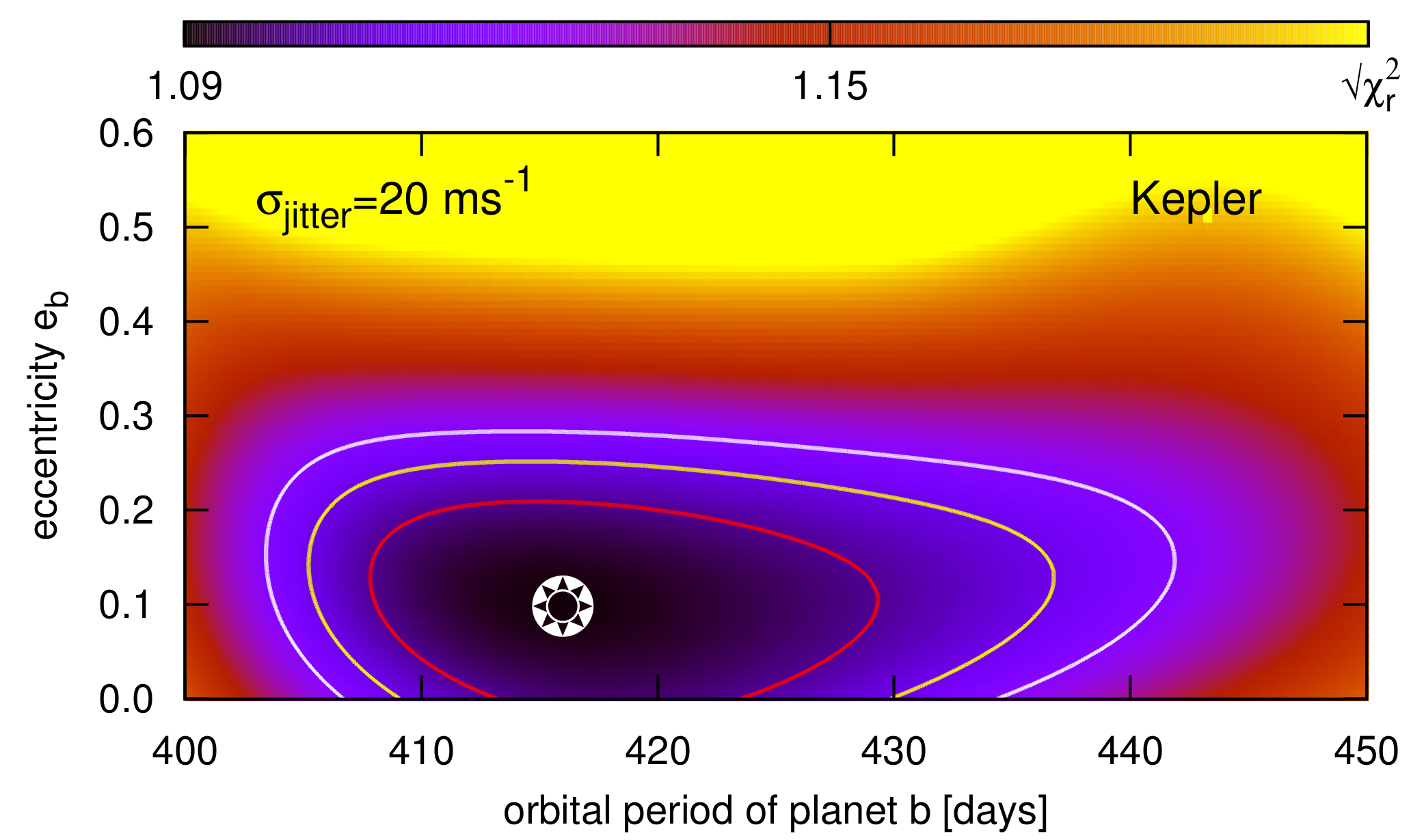} 
   \includegraphics[width=0.42\textwidth]{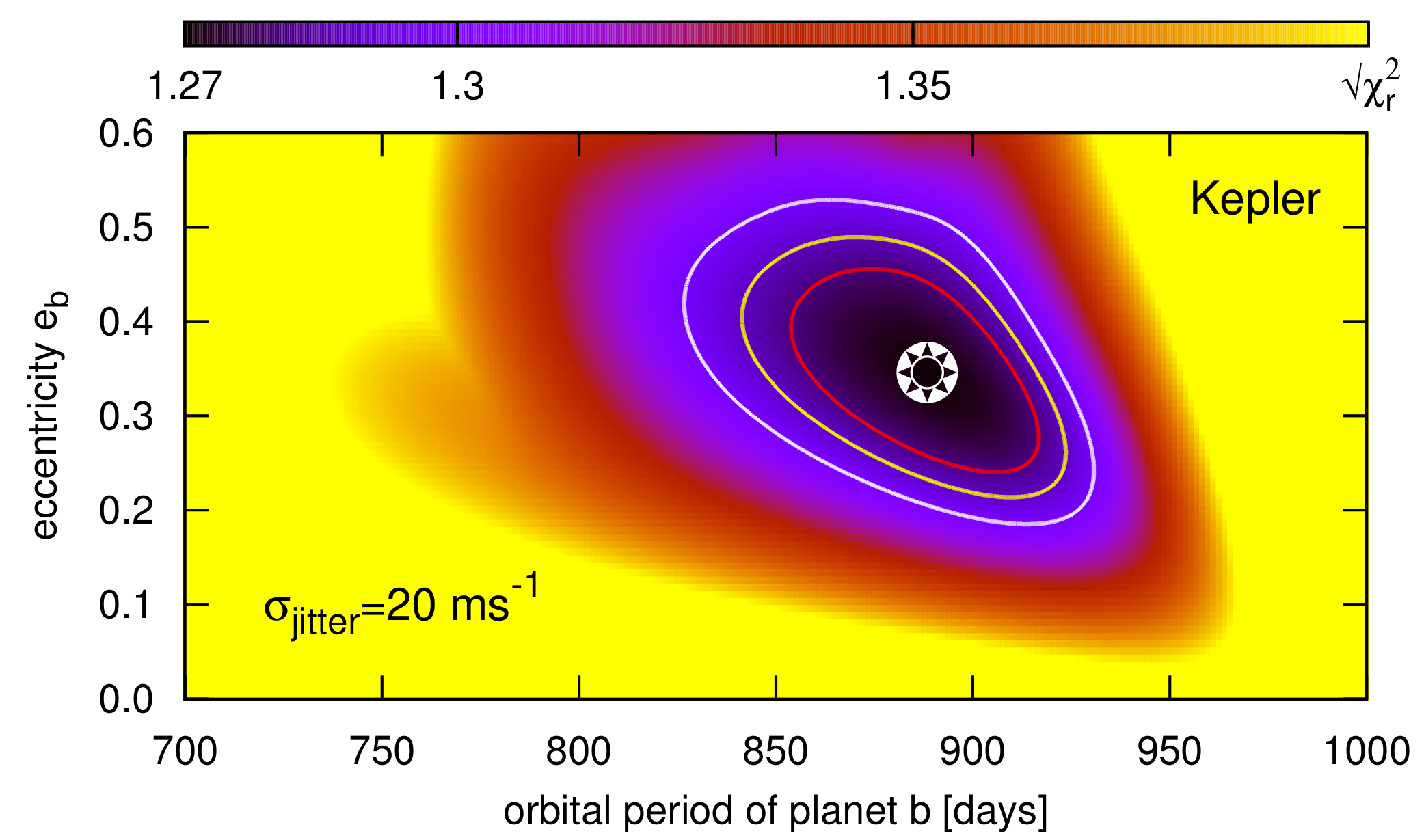} 
 }
}
}
\caption{
The topology of $\Chi$ in the orbital period--eccentricity plane of the
 putative planet for two values of the jitter uncertainty. The top row is for
 $\sigma_{\idm{jitter}}=5$~\mvs, the bottom plot is for
 $\sigma_{\idm{jitter}}=20$~\mvs. Smooth, coloured curves are for formal
 1,2,3$\sigma$ levels of the best-fit model marked with a star symbol. 
}
\label{fig:fig5}
\end{figure*}

The results are illustrated in Fig.~\ref{fig:fig5}, in the form of $\Chi$
parameter scans illustrated in the orbital period $P_{\idm{p}}$ --
eccentricity $e_{\idm{p}}$ plane of the planet. Curiously, for both selected
jitter uncertainties, we found \corr{a} well defined secondary minimum for
$P_{\idm{p}}\sim 900$~days and $e_{\idm{p}} \sim 0.3$, 
\corr{besides} the best-fit
model reported by \cite{ramm2009} with $P_{\idm{p}}\sim 420$~days and
$e_{\idm{p}} \sim 0.1$. A choice of the jitter uncertainty may lead to
significant changes of the $\Chi$ shape and alters formal errors of the
best-fit parameters. This effect is important for fixing parameter bounds in
dynamical studies made after the best-fit models are found. Moreover, because
the secondary best-fit configuration has the planetary period comparable with
the binary period, its interpretation in terms of the osculating elements
becomes even more questionable than in the case of the primary minimum of
$\Chi$, as underlined in the introduction. The secondary solution might
correspond to 1:1~MMR \corr{which obviously would require a more general} $N$-body model of the
RV which takes into account all mutual interactions between the bodies in the
system \citep{Laughlin2001}.
%
%
\subsection{The self-consistent, Newtonian model of the RV}
%
As we \corr{have} shown above, the Keplerian model of \nuoct{} data is non-unique,
permitting two different solutions. It is also well known  that it is
independent on orbital inclinations and true masses. Due to the massive
secondary, the RV-model including mutual interactions between all system
components might constrain these parameters, in spite of a narrow
observational window, like in other interacting extrasolar planetary systems
\citep{Laughlin2001}. Hence, in the second fitting experiment, to compute the
RVs of the primary, we integrated the $N$-body equations of motion. In this
model, the primary mass $m_{\idm{bin}}$, its semi--major axis $a_{\idm{bin}}$,
eccentricity $e_{\idm{bin}}$ of the binary, pericenter argument
$\omega_{\idm{bin}}$, inclination $i_{\idm{bin}}$ and the initial mean anomaly
${\cal M}_{\idm{bin}}$ of the binary were allowed to vary within the formal
$3\sigma$ error bounds, in accord with the combined RV and astrometric
solution in \cite{ramm2009}. The nodal longitude of the binary was fixed,
$\Omega_{\idm{bin}}=87^{\circ}$. The formal error of this element is
irrelevant for the dynamics of the system, because selecting the nodal line of
the binary corresponds to a choice of the $x$--axis of the reference frame.
The planet mass $m_{\idm{p}}$, its inclination $i_{\idm{p}}$, the nodal
longitude $\Omega_{\idm{p}}$ and the remaining orbital elements with the RV
offset were added to the set of 14~free parameters of the fit model. The
uncertainties of the measurements in \citep{ramm2009} were rescaled in
quadrature, as in the Keplerian model, to account for the intrinsic stellar RV
variability. 

Similar to the Keplerian model, we explored the parameter space with the
hybrid algorithm. After an extensive search, we selected the best-fit $N$-body
configuration. Then we reconstructed the shape of $\Chi$ around this nominal
solution in 2-dimensional planes of various orbital elements. Its parameters
derived with $\sigma_{\idm{jitter}}=5$~\mvs are given in Table~1 (Fit I). The
RV data in \cite{ramm2009} consist of many ``clumps'' of measurements done
during one night, which cover less than 0.1~day. Such data points were binned,
and the resulting data set consists of 91~measurements. This step was helpful
to improve the CPU-efficiency, due to necessity of computing numerical
derivatives in the LM algorithm. We examined that $\Chi$-scans can be
reproduced with the full, original data set. \corr{Obtaining a smooth
shapes of $\Chi$ for non-binned data is difficult, due to numerical errors
introduced by the numerical derivatives}.

\begin{figure*}
\centerline{
\vbox{
 \hbox{
   \includegraphics[width=0.42\textwidth]{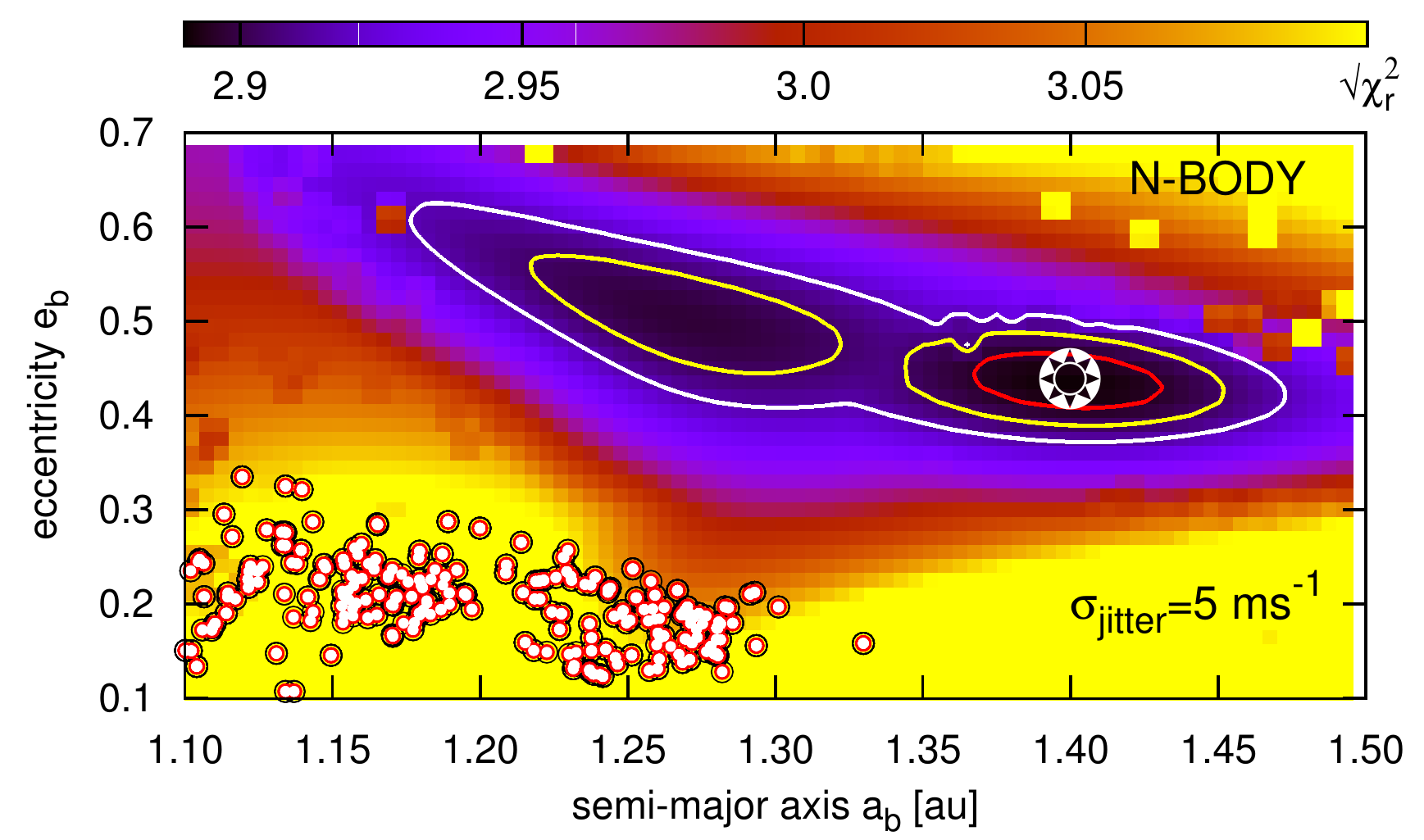} 
   \includegraphics[width=0.42\textwidth]{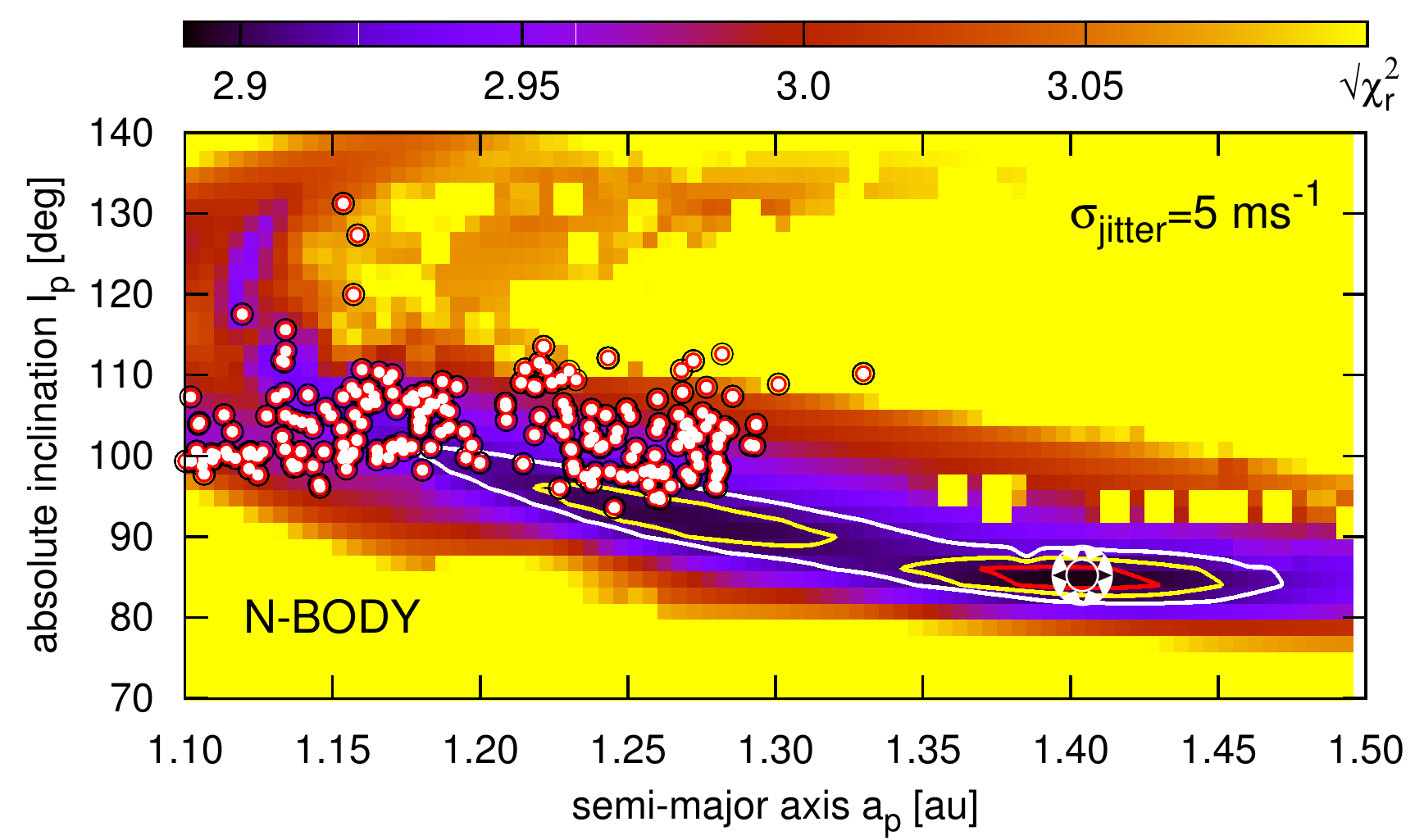} 
 }
 \hbox{
   \includegraphics[width=0.42\textwidth]{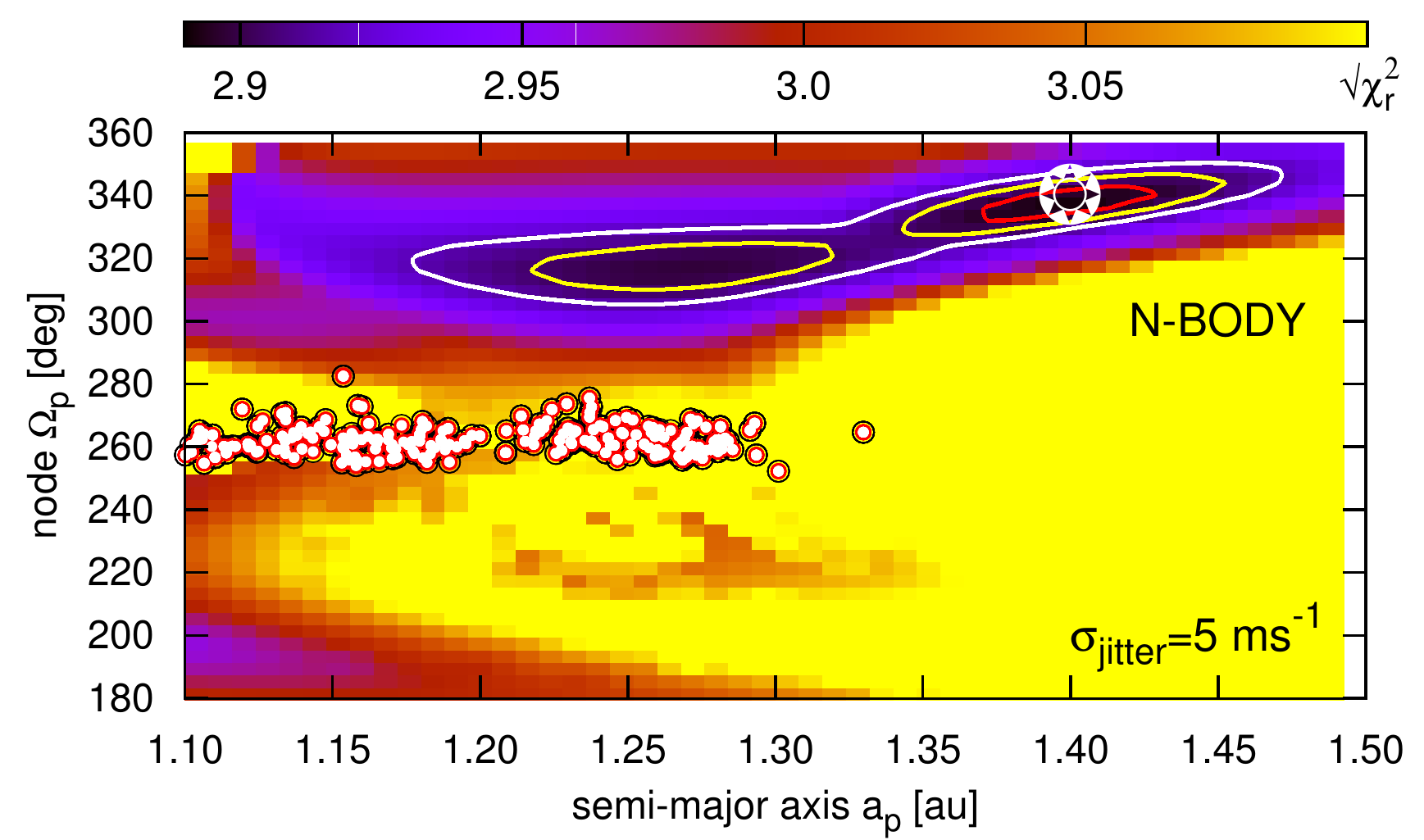}
   \includegraphics[width=0.42\textwidth]{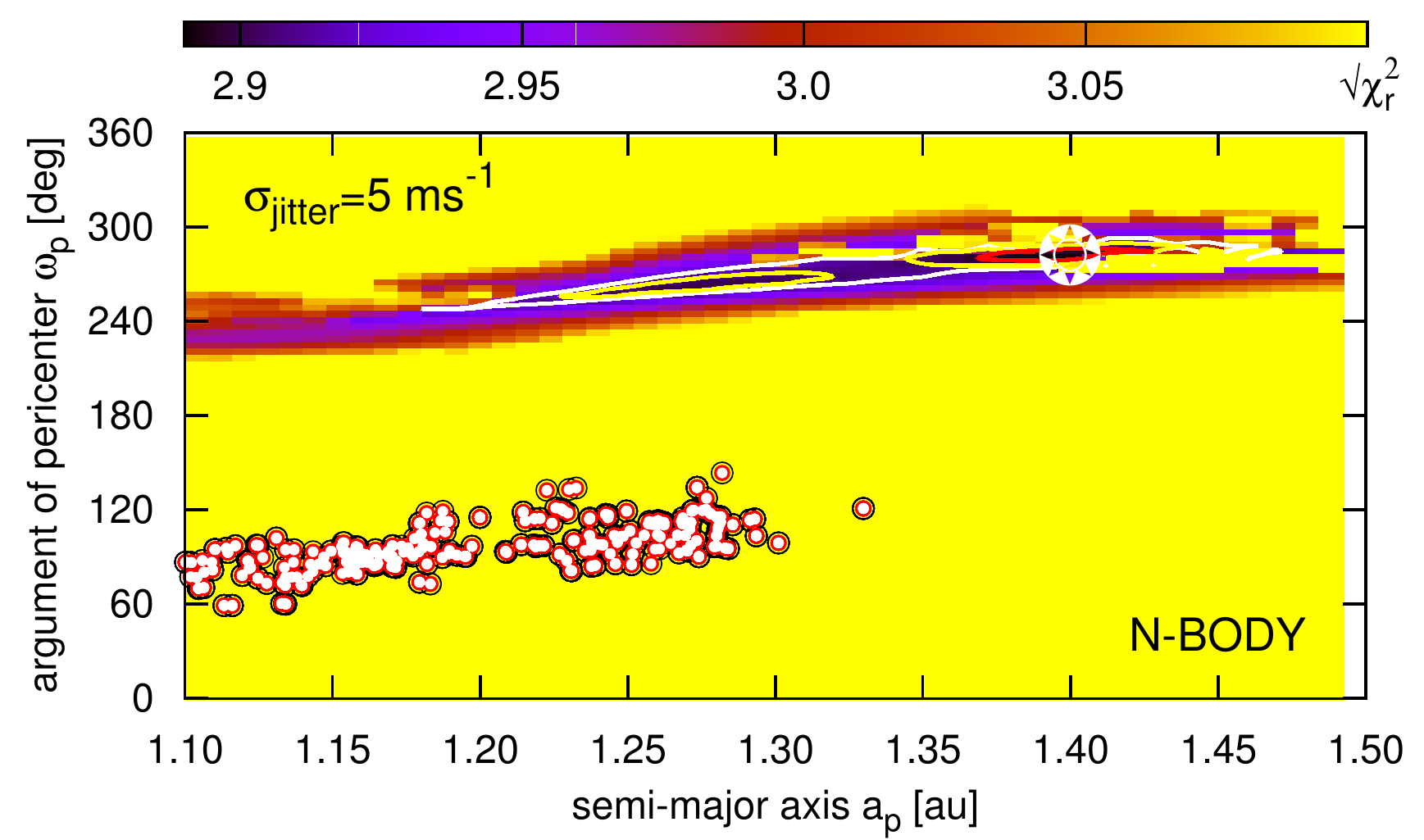}
 }
}
}
\caption{
The shape of $\Chi$ in different orbital planes of the planet computed in
terms of the self-consistent $N$-body model of the RV. Jitter uncertainty is
$\sigma_{\idm{jitter}} \sim 5$~\mvs{}. The RV data in (Ramm et al. 2012) are
binned (see the text for details). Quality of solutions is color coded.
Smooth, coloured curves are for formal 1,2,3$\sigma$ confidence levels of the
best-fit model marked with an asterisk. Its elements recomputed on the
original data set are given in Table~1 (Fit~I). Smaller, filled circles mark
{\em stable} solutions found with through the GAMP search, and projected onto
planes of orbital osculating elements at the epoch of the first observation in
(Ramm et al. 2009). An rms of these solutions is less than 28~\mvs{}. 
}
\label{fig:fig6}
\end{figure*}

\begin{figure*}
\centerline{
\vbox{
 \hbox{
   \includegraphics[width=0.42\textwidth]{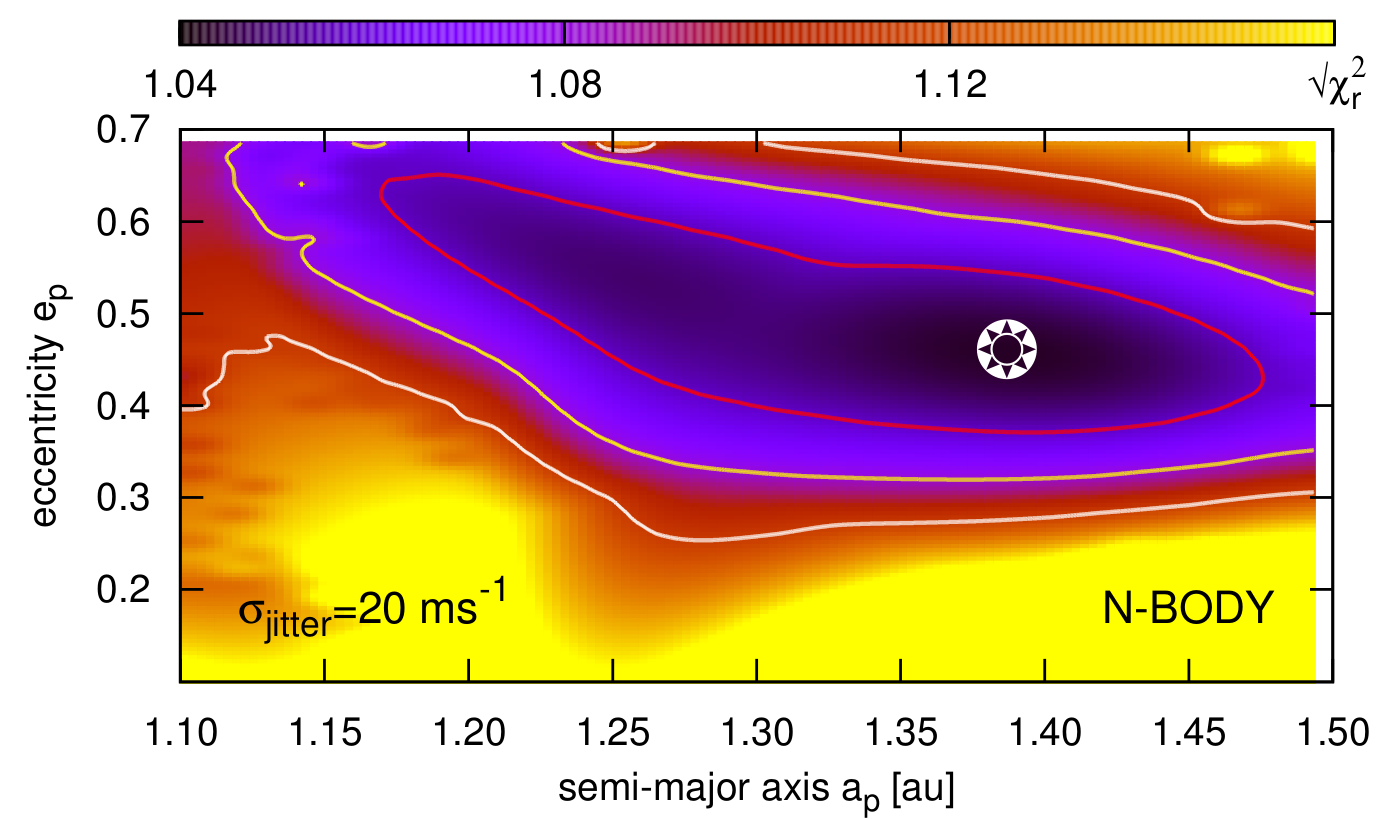}
   \includegraphics[width=0.42\textwidth]{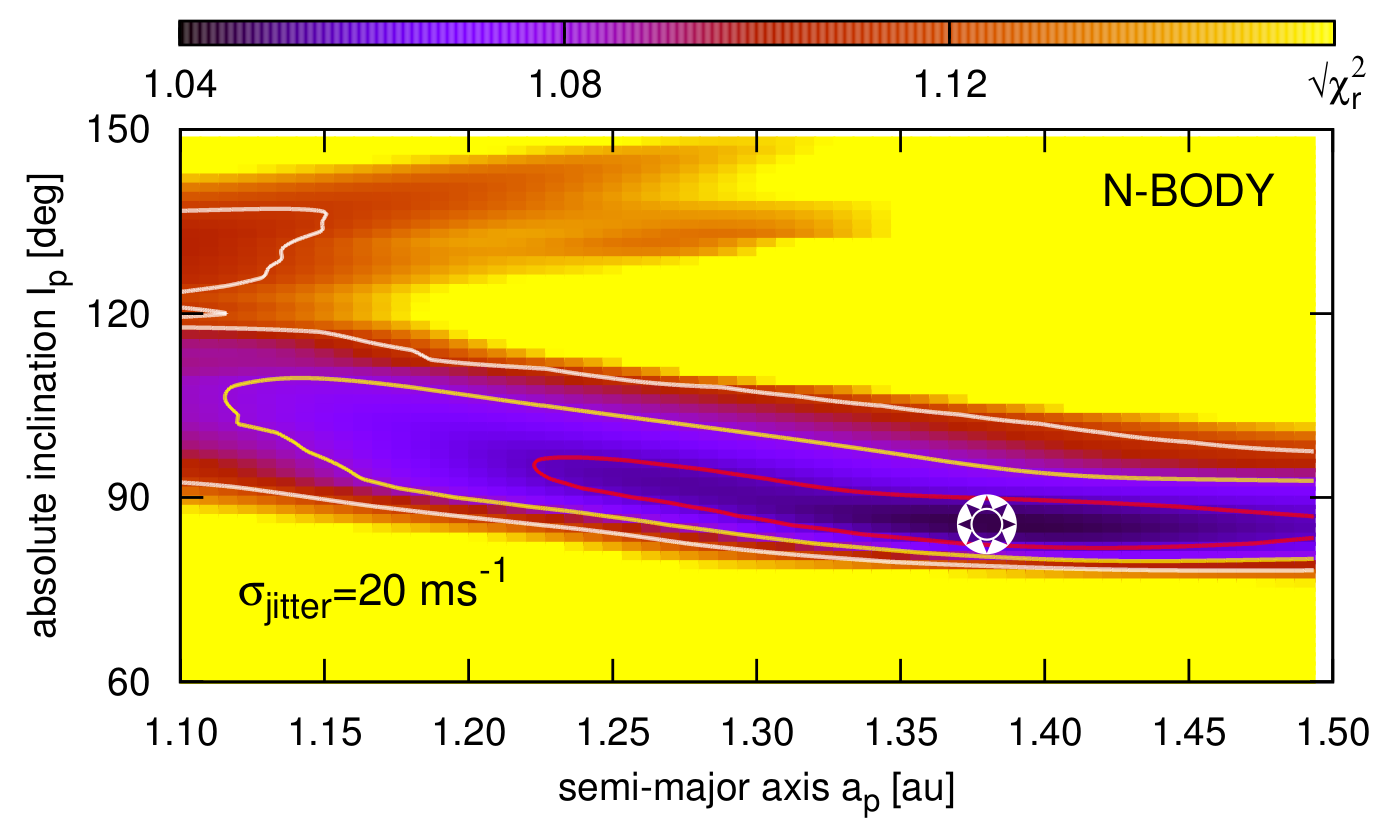}
 }
 \hbox{
   \includegraphics[width=0.42\textwidth]{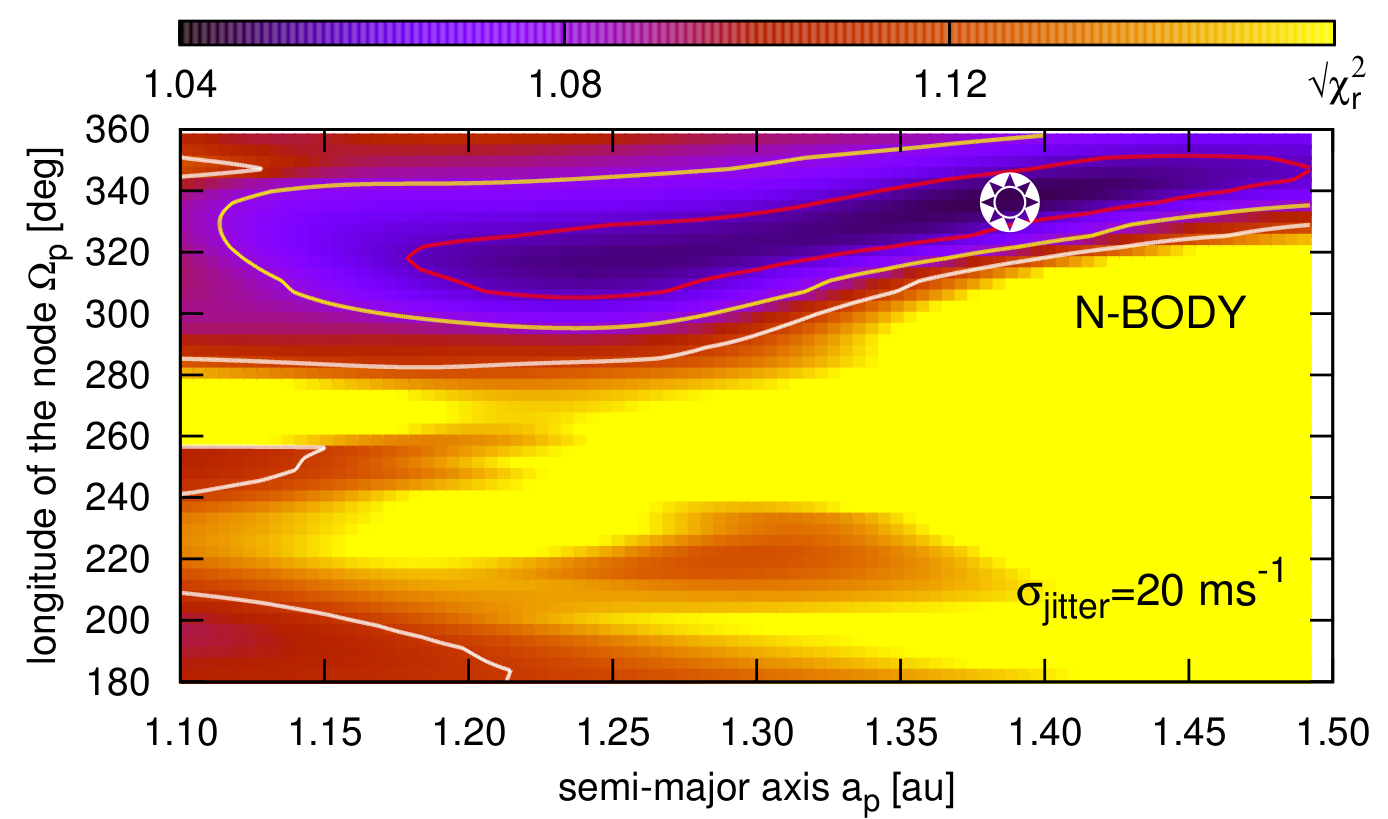}
   \includegraphics[width=0.42\textwidth]{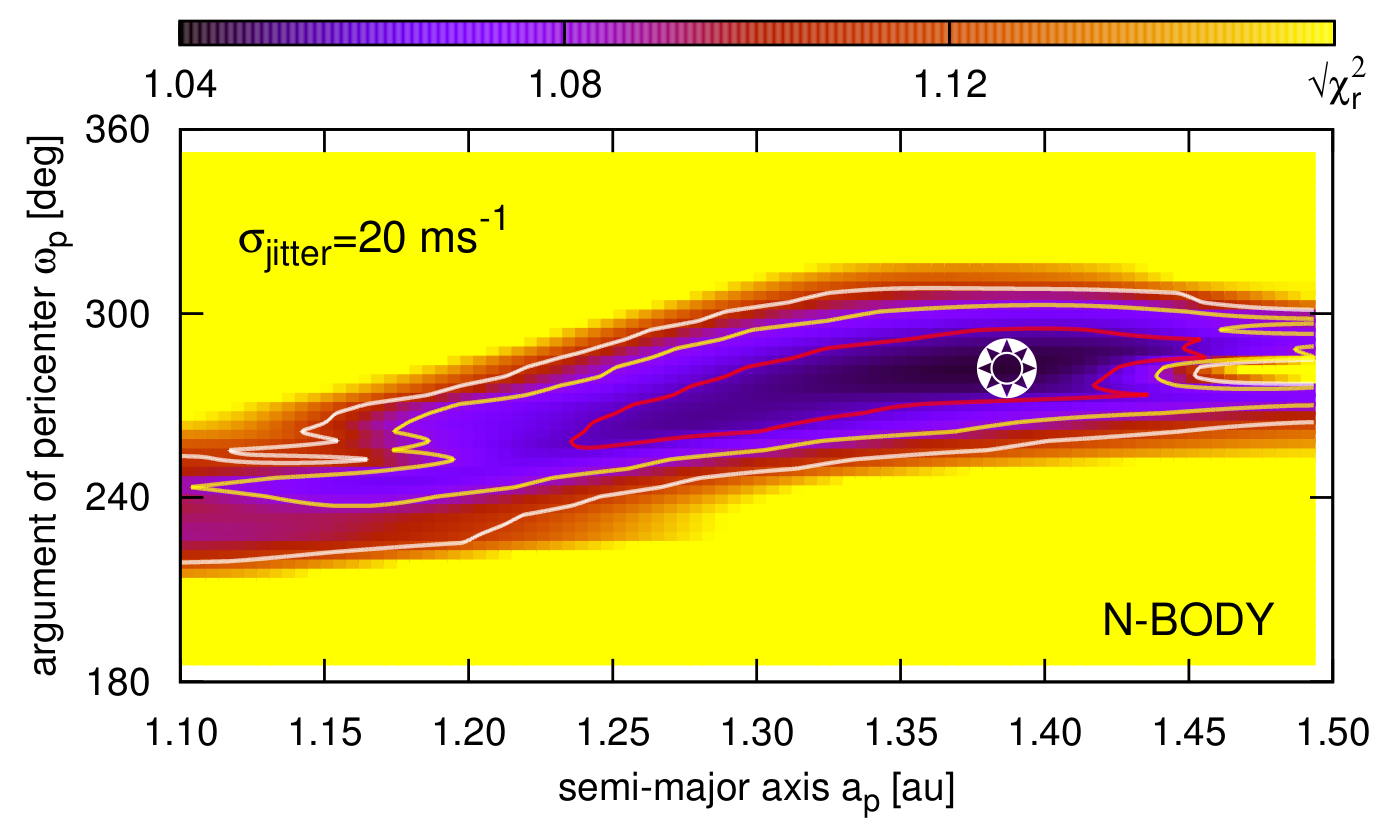}
 }
}
}
\caption{
The shape of $\Chi$ in terms of the Newtonian model illustrated in different
orbital parameter planes, as derived for  jitter uncertainty
$\sigma_{\idm{jitter}} \sim 20$~\mvs{}. Quality of solutions is color coded.
The best-fit configuration is marked with an asterisk.
}
\label{fig:fig7}
\end{figure*}

The $(a_{\idm{p}},e_{\idm{p}})$--plane (Fig.~\ref{fig:fig6}, the top-lef
panel) makes it possible to compare the topology of $\Chi$ in both RV-models.
Clearly, the best-fit Newtonian configuration is qualitatively different from
the Keplerian models found in the previous section \corr{which 
confirms our
earlier predictions}. The planet eccentricity may be as large as $\sim 0.45$.
The low estimate of $\sigma_{\idm{jitter}} \sim 5$~\mvs{} implies {\em two},
statistically distinct close minima of $\Chi$, for $a_{\idm{p}} \sim 1.25$~au, and better
model for $a_{\idm{p}} \sim 1.4$~au. These minima are resolved at the formal
$2\sigma$--error level of $\Chi$. If the jitter estimate is larger, likely
much more realistic, only one solution persist (see Fig.~\ref{fig:fig7}, the
top left panel). In this case, the formal $1\sigma$ error of $a_{\idm{p}}$
spans almost $0.3$~au.

Even more interesting results are illustrated in the next panels of
Figs.~\ref{fig:fig6} and~\ref{fig:fig7} showing scans of $\Chi$ in the
semi-major axis--inclination, ---longitude of the node and ---the pericenter
argument planes, respectively. $\Chi$ scans computed for
$\sigma_{\idm{jitter}} \sim 5$~\mvs{} reveal two minima seen in
Fig.~\ref{fig:fig6}. Such scans for $\sigma_{\idm{jitter}} \sim 20$~\mvs{}
(Fig.~\ref{fig:fig7}) illustrate that the global minimum around
$(a_{\idm{p}}=1.40$~au, $e_{\idm{p}}=0.45)$ is well defined in all planes of
angular elements. This means that even a narrow observational window, covering
roughly 2 planetary periods, makes it possible to estimate {\em all angles} of
the orbit due to strong mutual interactions in the system. This experiment is
a convincing illustration that the Keplerian, kinematic model of \nuoct{} is
questionable because it looses information of the node and inclination of the
orbit. A proper estimate of the jitter uncertainty is vital to derive true
errors of the best-fit model.
We found that the best-fit model (Fit~I in Table~1) tends towards the lower
bound of inclination (69$^{\circ}$), with simultaneous increase of the
secondary mass. This effect appears because the inclination of the binary
cannot be constrained by the RV data alone. Likely, dynamically consistent
$N$--body model of the \nuoct{} system should also incorporate the \chip{}
astrometry, which would be helpful to fix the orbit of the binary fully
consistent with {\em all data} and the three-body model. We postpone this
analysis to another work. 
%
%
\subsection{Newtonian model with stability constraints}
\begin{figure*}
\centerline{
\hbox{
  \includegraphics[width=00.42\textwidth]{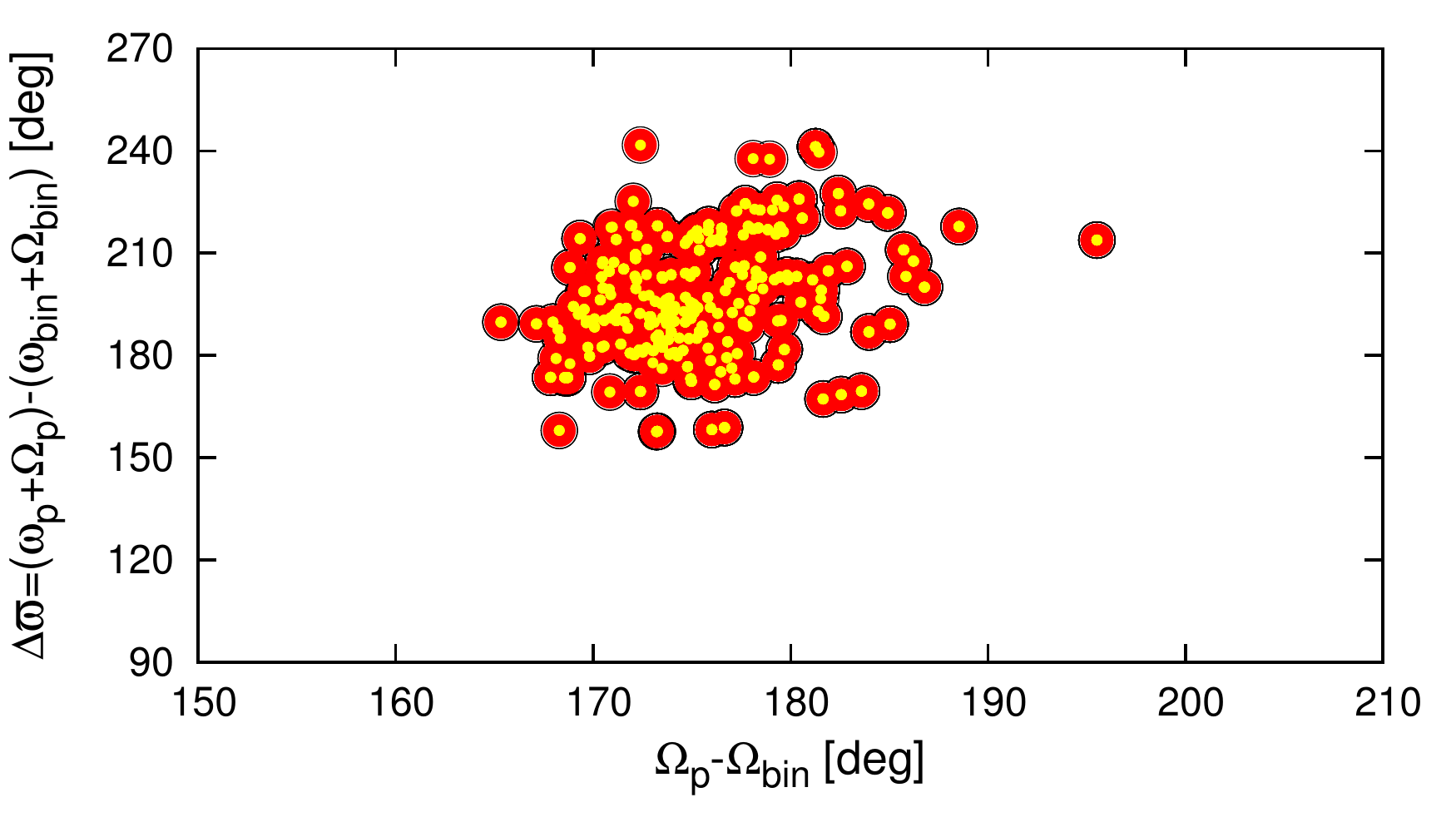}
  \includegraphics[width=00.42\textwidth]{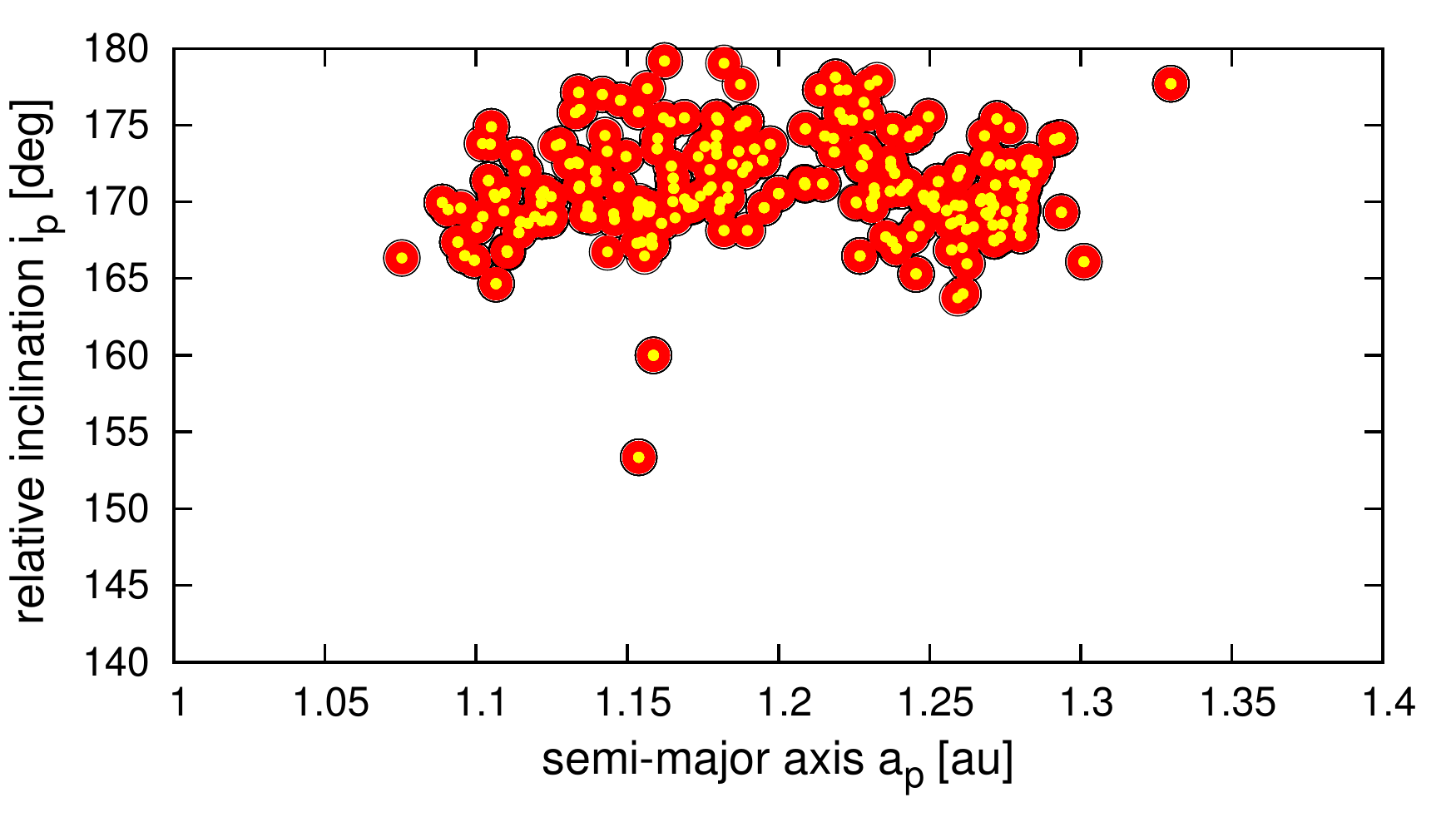}
}
}
\caption{
The statistics of stable best-fit models (see Fig.~6) projected onto selected
planes of the orbital osculating elements at the epoch of the first
observation in (Ramm et al. 2009). These fits are gathered in the hybrid GAMP
search (see the text for details). An rms of these solutions is less than
28~\mvs{}. 
}
\label{fig:fig8}
\end{figure*}

%
Unfortunately, the best-fit Newtonian model (Fit~I in Table~1) found through
the hybrid search leads to strongly unstable system. Due to relatively large
parameter error, stable configurations might be still possible in its
neighborhood. Hence, we conducted an additional, extensive search for {\em
stable} models imposing stability constraints \citep[GAMP
algorithm,][]{Gozdziewski2008}. In this approach, unstable models are
``penalized'' by an artificial, large value of $\Chi$ (or an rms) and are
excluded during the optimization. The results are shown in
Fig.~\ref{fig:fig6}. Stable fits exhibit systematically larger $\Chi$ than the
best--fit model in Table~1. To illustrate the statistics of such
configurations, we over-plotted solutions with an rms $<28$~\mvs{} (white,
filled circles) at the top of selected parameter scans of $\Chi$
(Fig.~\ref{fig:fig6}). Let us note that stable models have the MEGNO signature
$\sim 2$ for $2000~P_{\idm{bin}}$. This relatively short integration period
requires an acceptable CPU-overhead from the hybrid algorithm, still providing
at least 10-100 longer Lagrange stability time. Such solutions should survive
for $10^4$--$10^5$~$P_{\idm{bin}}$ or even longer integration time
\citep{Cincotta2003}.
The statistics in Fig.~\ref{fig:fig6} reveals that the formal best-fit
configurations in terms of ``pure'' $N$-body model are systematically shifted
by more than $3\sigma$ from the best-fit GAMP (stable) models. This regards
all orbital parameters, and in particular, the eccentricity and the argument
of pericenter. This large discrepancy may indicate that the 2-planet model is
in fact not proper, and should be rejected at all or modified. Curiously, we
found stable solutions {\em only} as retrograde orbits, actually in accord
with the original hypothesis in \citep{ec2010} 
\corr{and later investigated by
\cite{Quarles2012}}. This is illustrated in
Fig.~\ref{fig:fig8} showing that differences of the longitudes of periastrons
and of the nodal longitudes are well bounded around $180^{\circ}$, resulting
in apsidal lines of the orbits which are significantly {\em anti-aligned} (the
left panel). Large relative inclinations of these stable solutions (the right
panel) mean retrograde orbits in the range of $a_{\idm{p}}\sim$ around 1.2~au.

\begin{figure*}
\centerline{
 \vbox{
   \includegraphics[width=0.42\textwidth]{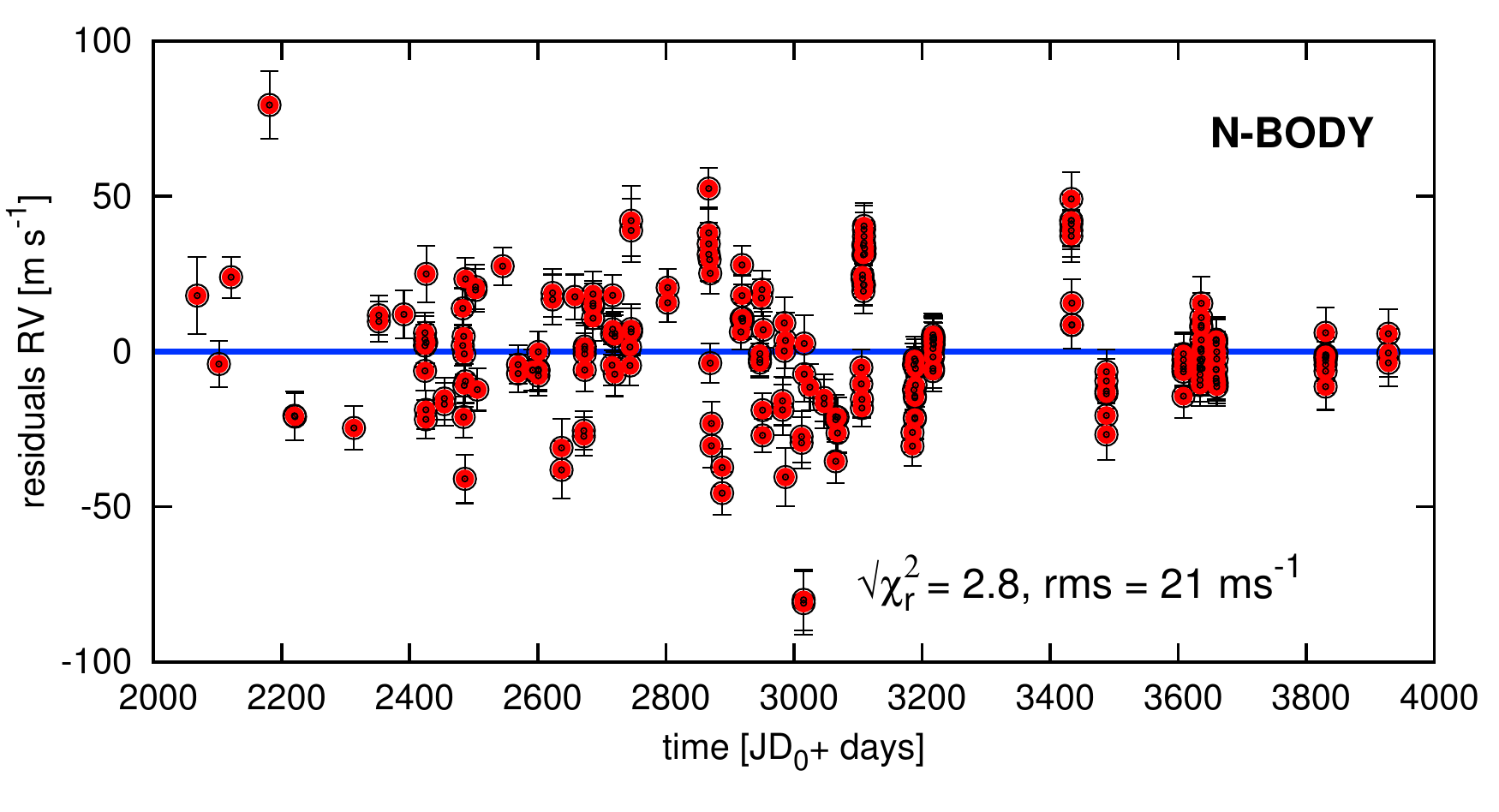} 
   \includegraphics[width=0.42\textwidth]{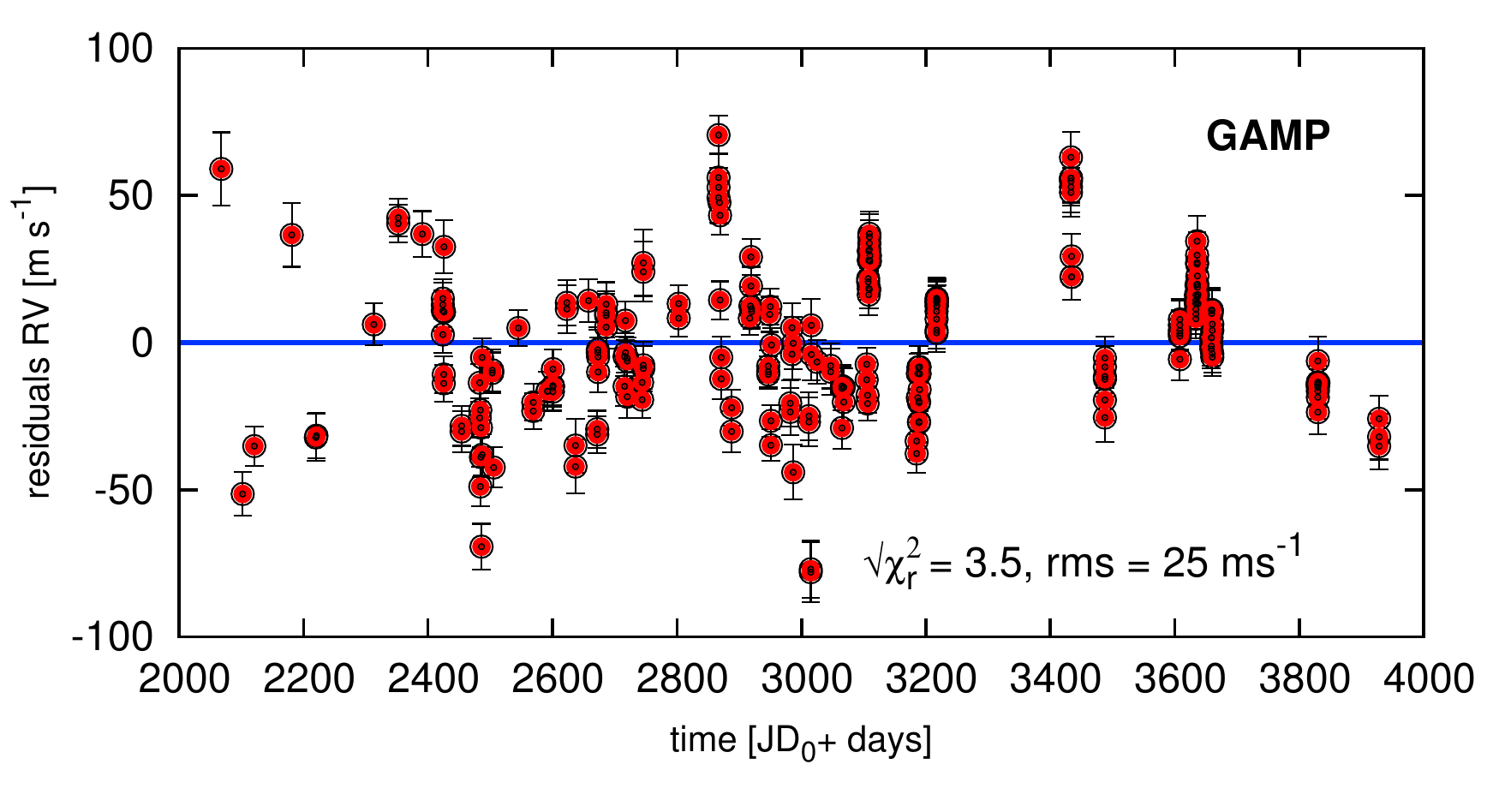} 
 }
}
\caption{
Residuals to representative self-consistent, Newtonian best-fit models illustrated in Fig.~\ref
{fig:fig8}. {\em The left panel} is for the formal, mathematical 
$N$-body solution (Fit~I in Table~1). {\em
The right panel} is for a stable, low-rms GAMP--derived model (Fit II in
Table~1).
}
\label{fig:fig9}
\end{figure*}

To compare the quality of these two families of models in terms of an rms, the
RV residuals of two selected best--fit models are displayed in
Fig.~\ref{fig:fig9}: the first one derived without stability constraints
(Fit~I in Tab.~1, \corr{the left} panel), and an low rms stable model
(\corr{the right} panel, Fit~II in Table~1). In both cases, the residuals exhibit a large
magnitude of $\sim 120$--$150$~\mvs, comparable with the semi-amplitude of the
RV signal $\sim 100$~\mvs{} due to the putative planet itself. In this sense,
stable solutions might be still acceptable, although their rms is
significantly larger from the rms of the formal best-fit model configuration.
A large variability of the RV residuals cannot be well explained by the
1-planet model. This may indicate strong stellar activity mimicking planetary
signal.

\begin{table*}
\label{tab:tab1}
\caption{
The best-fit astro-centric, osculating Keplerian elements of the \nuoctantis{}
planetary system at the first epoch in (Ramm et al. 2009). The RV data errors
in the discovery paper are rescaled with the jitter uncertainty
$\sigma_{\idm{jitter}}=5$~\mvs in quadrature. The nodal longitude of the
binary orbit is fixed to 87~deg. Mass of the primary \nuoct{}~A is
1.4~$m_{\sun}$. Parameters of stable solutions (Fits II and III) are quoted
with many digits making it possible to demonstrate stability of these
solutions in very narrow stable zones. Formal errors of these fits can be
estimated as equal or similar to the uncertainties quoted for parameters of
formal, $N$-body Fit~I.
}
\centering
\begin{tabular}{|l|rr|rr|rr|}
\hline
& \multicolumn{2}{c}{Fit I ($N$-body, unstable)} & \multicolumn{2}{c}{Fit II
(GAMP, stable)} & \multicolumn{2}{c}{Fit III (GAMP, stable)}
\\
Parameter 
& {planet~\nuoct{}Ab} & {\nuoct{}~B} & {planet~\nuoct{}Ab} 
& {\nuoct{}~B} & {planet~\nuoct{}Ab} & {\nuoct{}~B}
\\
\hline
mass $m$ [m$_{\idm{Jup}}$]
  & 3.54 $\pm$ 0.07   &  572 $\pm$ 15  &  1.92596   &  560.62366   &   2.135457  &  559.2707       \\
semi-major axis $a$ [AU]    
  & 1.41 $\pm$ 0.02   &  2.533 $\pm$ 0.005 & 1.16365    &  2.52813   &  1.160281   &   2.526478     \\
eccentricity $e$
  &  0.40 $\pm$ 0.01  &   0.2331 $\pm$ 0.0006 & 0.13139    & 0.23881    & 0.247587    & 0.238713       \\
inclination $i$ [deg]     
  &  84.9 $\pm$ 2.2   & 69.0 $\pm$ 3.3   &  110.21102   & 71.28090    &   102.4152  &   71.681      \\
longitude of the node $\Omega$ [deg] 
  & 289.2 $\pm$ 4.2   & 87.0 (fixed)      & 262.29811    &  87.0 (fixed)        &   260.0224  &  87.0 (fixed)      \\
argument of pericenter $\omega$ [deg]     
  & 340.6 $\pm$ 2.6  & 75.0 $\pm$ 0.1   &  127.26299   & 74.59137    &  88.86501   &  74.85992      \\
mean anomaly ${\cal M}(t_0)$ [deg]
  & 289.2 $\pm$ 4.2  & 339.5 $\pm$ 0.l   & 133.16327    &  339.762973   &  165.3685   &  339.5231       
\\
\hline
$\Chi$  
& \multicolumn{2}{c}{2.79} & \multicolumn{2}{c}{3.47} & \multicolumn{2}{c}{3.44} 
\\
RV offset $V_0$ [m s$^{-1}$]      
& \multicolumn{2}{c}{-6417 $\pm$ 1} & \multicolumn{2}{c}{-6473.682} & \multicolumn{2}{c}{-6475.317}
\\
rms~[m s$^{-1}$]  
       & \multicolumn{2}{c}{19.9} & \multicolumn{2}{c}{25.2}  &
\multicolumn{2}{c}{24.8}
\\
\hline
\end{tabular}
\end{table*}

\subsection{Stability of the best-fit GAMP configurations}
%
To illustrate the dynamical neighborhood of the best--fit GAMP models, we
selected two examples of low rms configurations. Their osculating elements at
the epoch of the first observation in \citep{ramm2009} are given in Table~1
(Fits II and III). Many digits are quoted, to reproduce elements of these fit
possibly exactly, to make it possible to reproduce their stable evolution due
to complex and chaotic neighborhoods. The formal errors may be estimated
graphically in Fig.~\ref{fig:fig6} or \corr{interpreted} as uncertainties of the formal
Fit~I. The primary mass is fixed as equal to 1.4~\ms. An rms of this
solution is $\sim25$~\mvs.

The top-left panel of Fig.~\ref{fig:fig10} illustrates the MEGNO dynamical map
in the $(a_{\idm{p}},i_{\idm{p}})$--plane computed for Fit~II (see Tab.1).
This map shows a global view of the phase space. Note that inclination
$i_{\idm{p}}$ is {\em absolute w.r.t. the astrometric frame}, in which the
binary inclination is bounded around $i_{\idm{bin}}\sim 71^{\circ}$. This figure
reveals that the phase space is mostly strongly chaotic. The best--fit
configuration is found in a narrow stability island close to the 7:2~MMR. For
a reference, other significant MMRs are labeled. A close-up of this map shown
in the top--right panel of Fig.~\ref{fig:fig10} reveals an extremely complex
structure of this regular island. It spans a tiny region covering $\Delta
a_{\idm{p}} \sim 0.02$~au and $\Delta i_{\idm{p}} \sim 20^{\circ}$. In this
small island, the phase space has again a sophisticated, fractal-like
structure of the Arnold web. 

The integration time used to derive the global dynamical map in the left panel of
Fig.~\ref{fig:fig10} is $1 \times 10^4 P_{\idm{bin}}$. Subsequent close-ups shown in
Fig.~\ref{fig:fig10} were integrated for $2 \times 10^4 P_{\idm{bin}}$ up to
$1 \times 10^5 P_{\idm{bin}}$ per each point, depending on the magnification factor. To
detect weak MMRs and their sub-resonances, such long integration times are
indispensable. The resolution of these maps is $1440\times 900$ pixels.
{Calculations were performed with the help of the \mechanic{} installed on the
UV~SGI supercomputer {\tt chimera} of the Pozna\'n Supercomputing Centre
(Poland)\footnote{Supercomputer {\tt chimera} has 2048 CPU cores, and runs
spanning up to 24~hrs occupied all available hardware resources. We did not
observed any bottlenecks in the master--worker inter-communication. This nice
performance of the \mechanic{} code was expected thanks to a small overhead of
the MPI data broadcasts. These computations validated the software and its
ability of stable long-runs on a few thousands of CPU cores.}.

We computed a similar set of dynamical maps for another stable model, with
larger eccentricity (Fit~III in Table~1). The results are illustrated in
Fig.~\ref{fig:fig11}. The top left panel illustrates the dynamical map in the
($a_{\idm{p}},e_{\idm{p}}$)--plane, showing a narrow stable island of this
solution, the rest of plots are for the ($a_{\idm{p}},i_{\idm{p}}$)--plane
and its close-ups.
\begin{figure*}
\centerline{
\vbox{
   \hbox{
   \hbox{\includegraphics[width=00.42\textwidth]{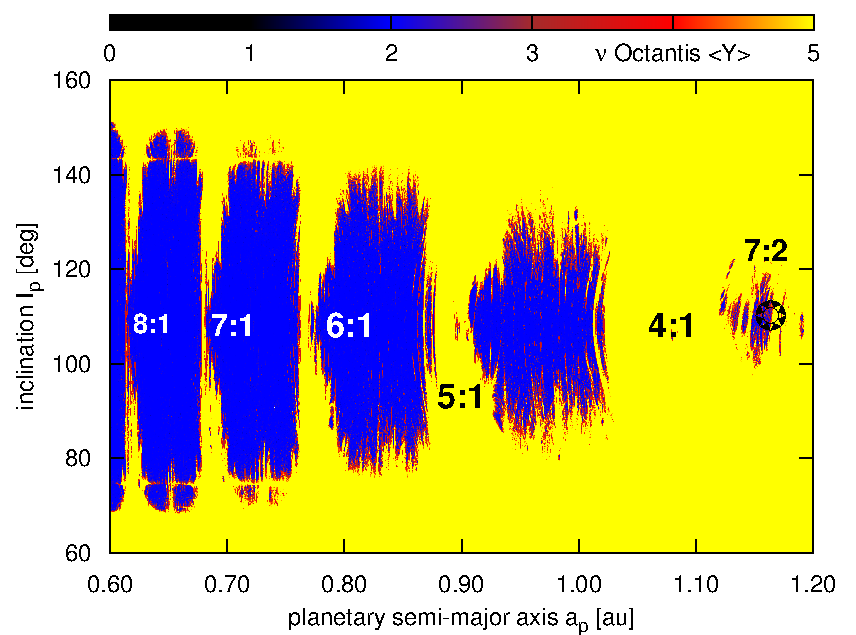}} 
   \hbox{\includegraphics[width=00.42\textwidth]{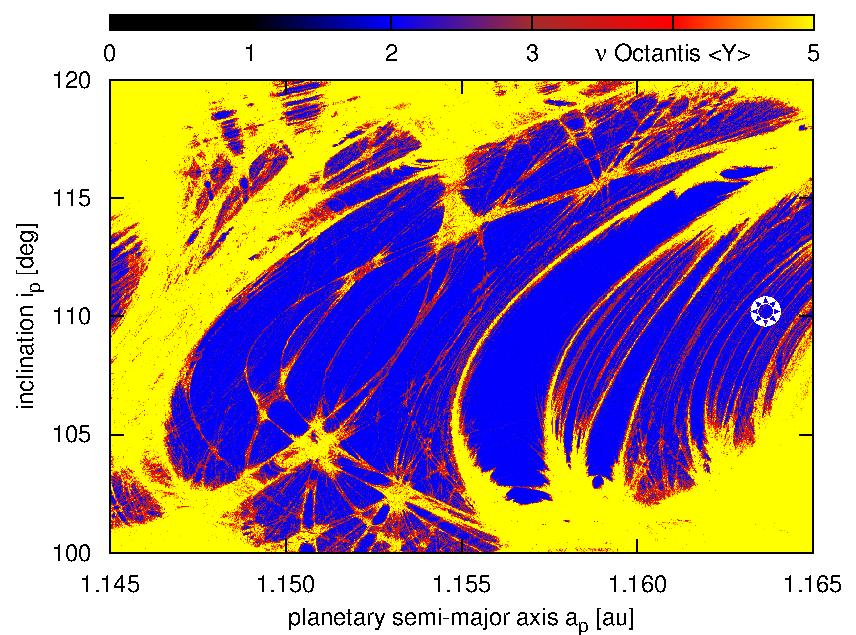}} 
   }
\hbox{
   \hbox{\includegraphics[width=00.42\textwidth]{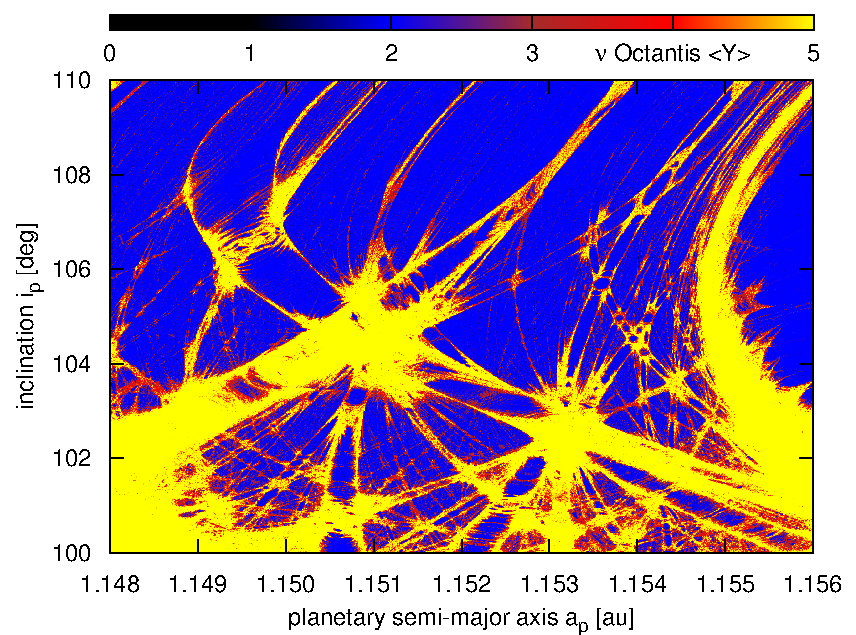}} 
   \hbox{\includegraphics[width=00.42\textwidth]{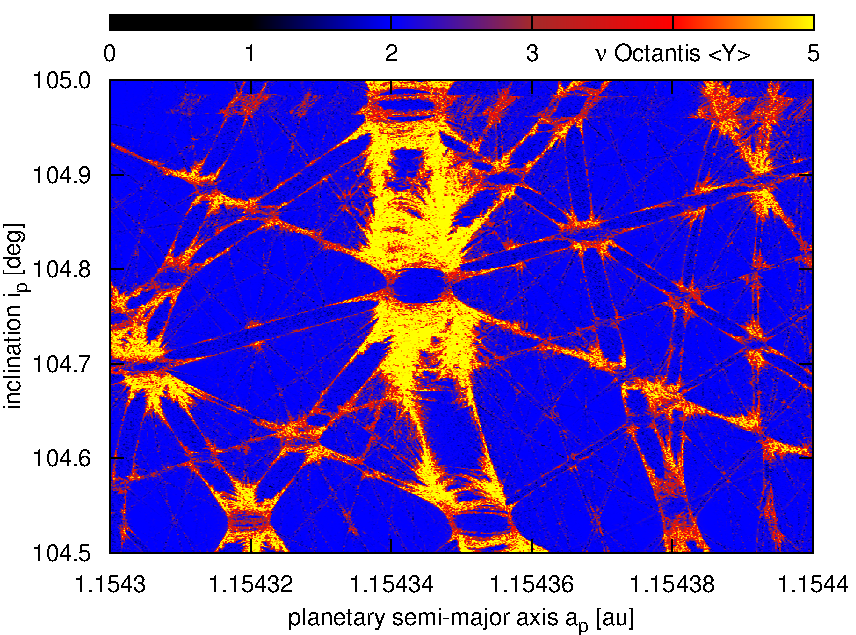}}  
   }
}
}
\caption{
Dynamical MEGNO maps of a stable model found in the GAMP search illustrated in
the ($a_{\idm{p}},i_{\idm{p}})$--plane. {\em The top left panel} is for the
global view of the phase space, {\em the top right} map is for stability
island of the solution. Subsequent maps are for close-ups of the neighborhood
of the best--fit solution. Osculating, {\em astrocentric} elements of the
planet and the secondary are given in Table~1 (Fit II). The nominal elements
are marked with the star symbol. The raw resolution of the maps is
1440$\times$900 data points.
}
\label{fig:fig10}
\end{figure*}
\begin{figure*}
\centerline{
\vbox{
   \hbox{
   \hbox{\includegraphics[width=00.42\textwidth]{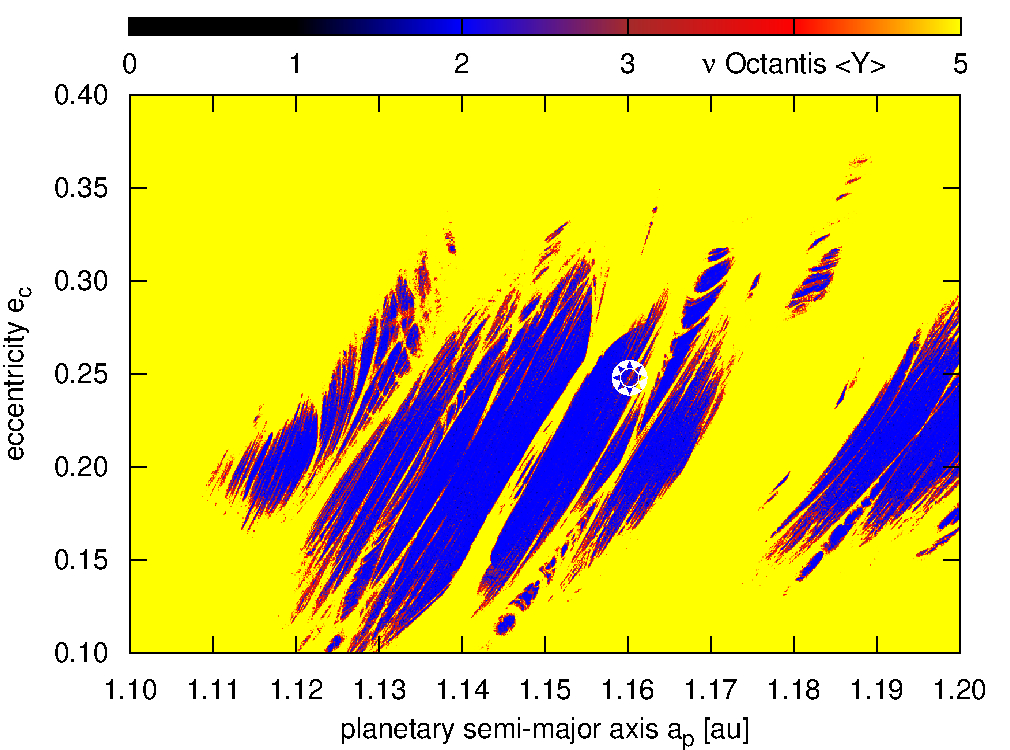}} 
   \hbox{\includegraphics[width=00.42\textwidth]{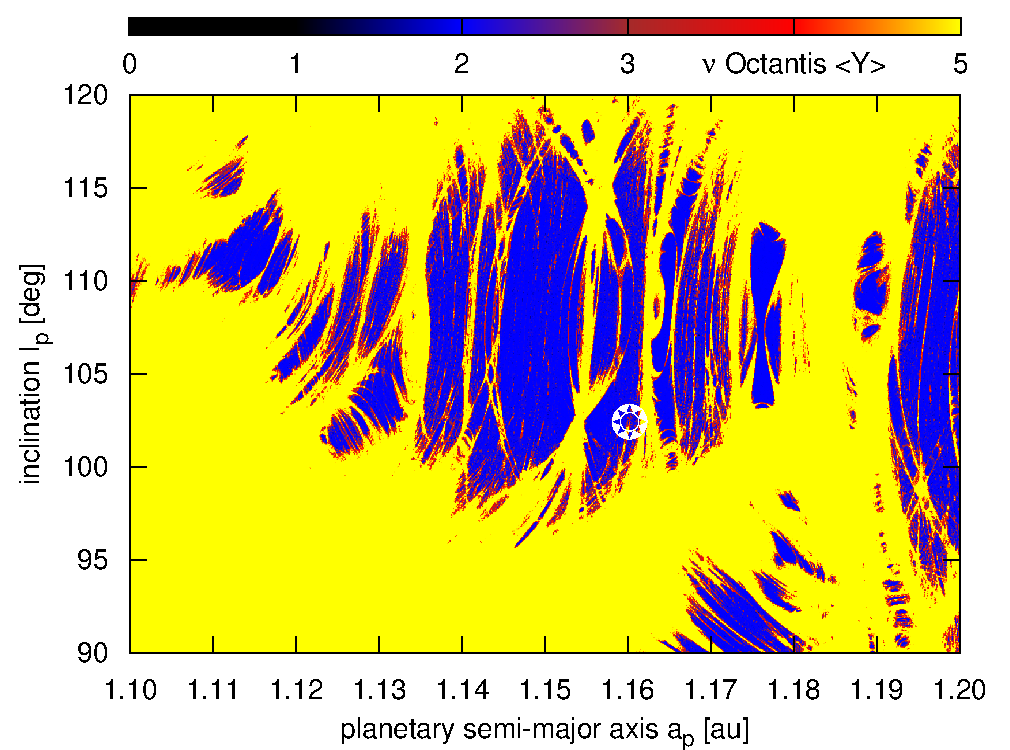}} 
   }
\hbox{
   \hbox{\includegraphics[width=00.42\textwidth]{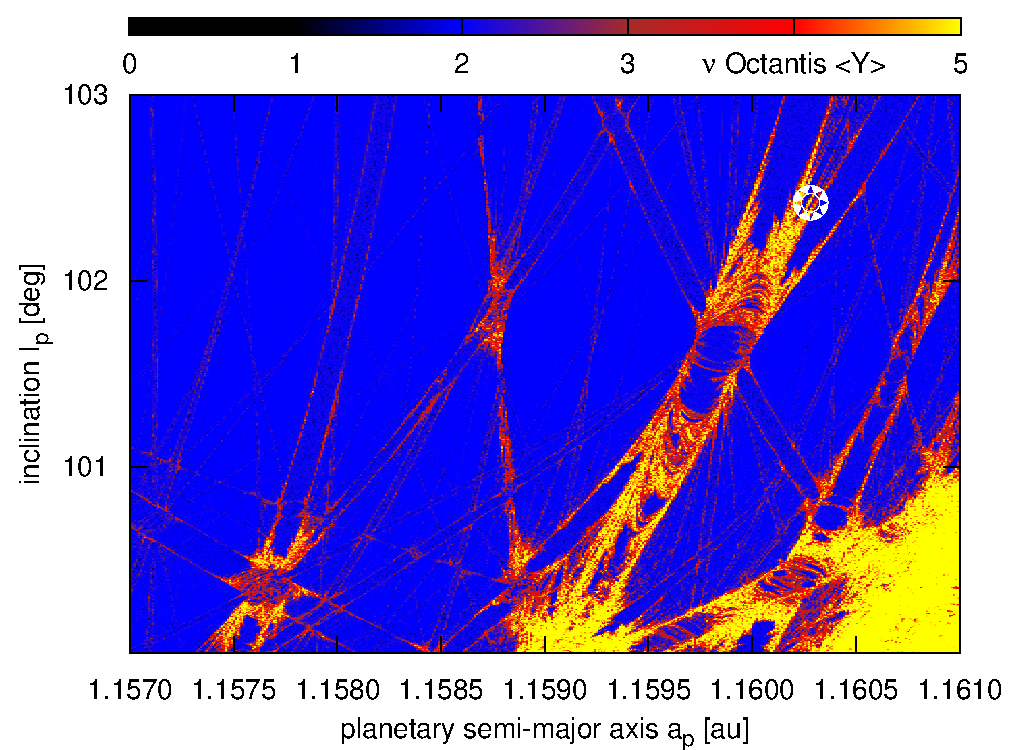}} 
   \hbox{\includegraphics[width=00.42\textwidth]{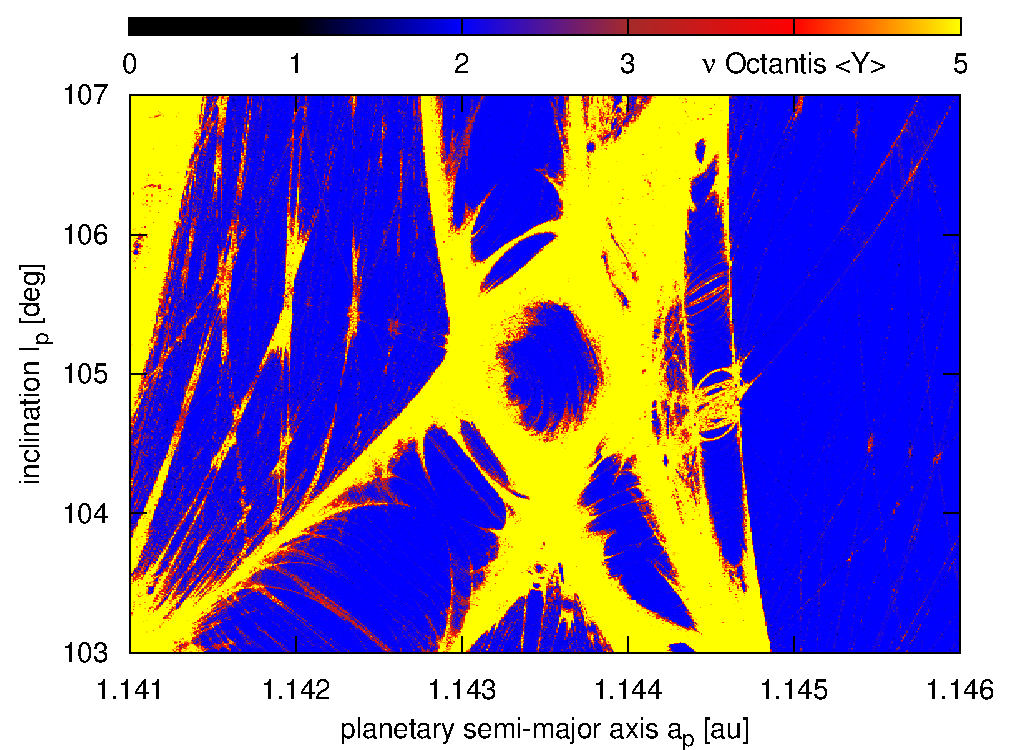}} 
   }
}
}
\caption{ 
Dynamical MEGNO maps of a stable model with moderate eccentricity of the
planet as found in the GAMP search illustrated in the
($a_{\idm{p}},e_{\idm{p}})$ and ($a_{\idm{p}},i_{\idm{p}})$--plane. Bottom
panels show close-ups of the top right panel. Osculating, {\em astrocentric}
elements of the planet and the secondary are given in Table~1 (Fit III). The
nominal elements are marked with the star symbol. The raw resolution of the
maps is 1440$\times$900 data points.
}
\label{fig:fig11}
\end{figure*}
To reveal the Arnold web in more detail, we computed a number of close-ups
shown in panels of Figs.~\ref{fig:fig10}--\ref{fig:fig12}. These maps reveal
many weak sub-resonances, which form a regular net present at any magnification
level. These structures are particularly well seen in the bottom-right panel
of Fig.~\ref {fig:fig12}. They closely resemble the Arnold web of the model
Hamiltonian \citep{Froeschle2000}. Likely, it has been never illustrated with
such a sharpness and details. 
We also tested other solutions from the GAMP--derived set. The results are
similar. The best--fit models which obey reasonably well the observational
constrains are confined to only narrow stability islands. 
\begin{figure}
\centerline{
\vbox{
   \centerline{\hbox{\includegraphics[width=0.50\textwidth]{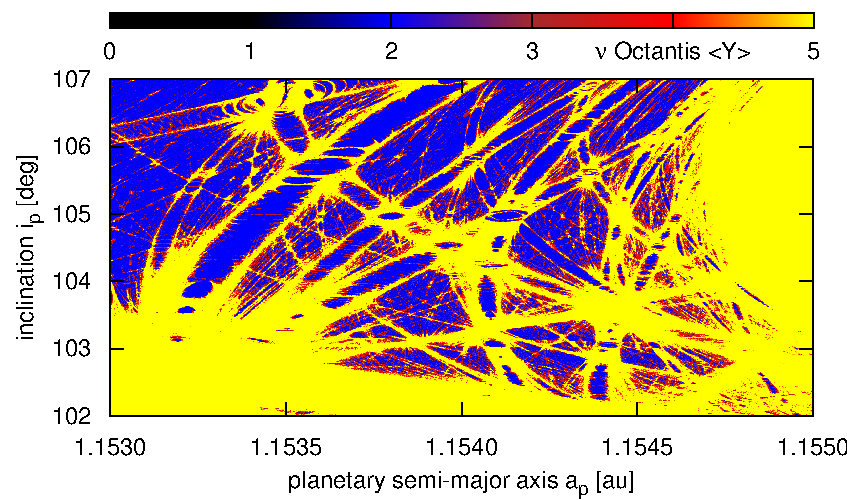}}}
}
}
\caption{
A close-up of a stability island of the best-fit, stable GAMP Fit~II (Table~1)
illustrated in Fig.~\ref{fig:fig10}. The resolution is $1440\times 900$
pixels. The integration time per pixel is $3 \times 10^5$ orbital periods of the binary. 
}
\label{fig:fig12}
\end{figure}
%
%
\section{Conclusions}
%
%
In this work, we aim to improve the original Keplerian models of the \nuoct{}
system. We re-analysed the RV observations in \citep{ramm2009}. The results
confirm our predictions on a significance of the mutual interactions between
the secondary and the putative planet. These interactions are so strong that
-- if the planet is real -- it might be possible to derive the orientation of
the orbit with respect to the binary, and its true mass, in spite of a narrow
observational window spanning less than 2 orbital periods.

We confirm that stable retrograde solutions in the \nuoct{} planetary system
\corr{are possible as
proposed in \citep{ec2010,Quarles2012}}. However, the observational constrains
imply such orbits with apsidal lines initially {\em anti-aligned} with the
apses of the binary. The long-term stable solutions may be found only in
relatively small islands of regular, quasi-periodic motions exhibiting a
complex structure of the Arnold web. Stable models must be significantly
shifted in the parameter space from the $N$-body and Keplerian best-fit
models. It means that the 2-planet, S-type configuration is most likely
inadequate to fit the available data set. It remains highly uncertain how a
massive, Jovian planet could be trapped in such tiny stable regions, or how it
might form in globally unstable dynamical environment. 
\corr{The}
observational RV signal is very noisy, and the best-fit models reveal a large
scatter of residuals having amplitude comparable with the RV signal itself.
These arguments make the hypothesis of Jovian planet in orbit around
\nuoct{}~A questionable. Actually, this in accord with the doubts raised in the
discovery paper \citep{ramm2009}. New observations are required to confirm or
withdraw that explanation of the observed RV residual signal, and to search
for alternate dynamical models of the detected variability.

An efficient analysis of the observations of extrasolar planetary systems
involves not only different observational techniques and data sources, but
also orbital optimization algorithms, combined with analytical and numerical
stability studies. In this work, we propose and apply our new MPI-based task
management system \mechanic{} to perform high-resolution dynamical mapping of
the phase space. This framework is devoted to numerical analysis of large
sets of initial conditions, like the long-term integrations of the equations
of motion. We work on other applications, like quasi-global optimization with
the Genetic Algorithms of different observational models of the extrasolar
planetary systems. The \mechanic{} code with detailed technical documentation
and sample modules is freely available at the project website \citep
{mechanic-online}. 
%
%
\section{Acknowledgments}
%
We thank \corr{the anonymous referees for their reviews} that improved this paper.
This work is supported by the Polish Ministry of Science and Higher Education
through grant {N/N203/402739}. Computations were conducted thanks
to the POWIEW project of the European Regional Development Fund in
Innovative Economy Programme POIG.02.03.00-00-018/08.
%
%
\bibliographystyle{mn2e}
\bibliography{ms}
%
\end{document}